\documentclass[12pt,preprint]{aastex}

\usepackage{color,graphicx}
\usepackage{amssymb,amsmath,eurosym,psfrag}
\usepackage{mathrsfs}
\usepackage{enumerate}
\usepackage{natbib}

\usepackage[pdftex,plainpages=false]{hyperref} 
\usepackage[all]{hypcap}

\newcommand{\figref}[1]{Figure \ref{#1}}
         
\newcommand{\Frac}[2]{\frac{\raisebox{-0.5ex}{\ensuremath{#1}}}{#2}}

\newcommand{\vect}[1]{\mathbf{#1}}

\newcommand{\Matrix}[1]{\mathbb{#1}}
\newcommand{\tothe}[1]{\ensuremath{^{\hbox{\scriptsize #1}}}}

\newcommand{\csum}{\ensuremath{\textsf{c}\hspace{-0.75em}\sum}}
\newcommand{\fnc}[1]{\textsf{#1}} 

\def\bigvar#1#2{{\hbox{$\left#1\vbox to #2{}\right. \n@space$}}}
\def\n@space{\nulldelimiterspace=0pt \m@th}
\def\m@th{\mathsurround=0pt }

\begin{document}
 \author{Lucas Tarr \& Dana Longcope}
 \affil{Department of Physics, Montana State University, Bozeman, Montana 59717}

 \title{Calculating energy storage due to topological changes in emerging active region NOAA AR 11112}

 \begin{abstract}
   The Minimum Current Corona (MCC) model provides a way to estimate stored coronal energy using the number of field lines connecting regions of positive and negative photospheric flux.  This information is quantified by the net flux connecting pairs of opposing regions in a connectivity matrix.  Changes in the coronal magnetic field, due to processes such as magnetic reconnection, manifest themselves as changes in the connectivity matrix.  However, the connectivity matrix will also change when flux sources emerge or submerge through the photosphere, as often happens in active regions.  We have developed an algorithm to estimate the changes in flux due to emergence and submergence of magnetic flux sources.  These estimated changes must be accounted for in order to quantify storage and release of magnetic energy in the corona.  To perform this calculation over extended periods of time, we must additionally have a consistently labeled connectivity matrix over the entire observational time span.  We have therefore developed an automated tracking algorithm to generate a consistent connectivity matrix as the photospheric source regions evolve over time.  We have applied this method to NOAA Active Region 11112, which underwent a GOES M--2.9 class flare around 19:00 on Oct.$16\tothe{th}$, 2010, and calculated a lower bound on the free magnetic energy buildup of $\sim 8.25 \times 10^{30}$ergs over 3 days.
 \end{abstract}
 
 \date{Draft: \today}
 
 \section{\label{sec:intro}Introduction}
 
 It is now widely believed that solar flares are powered by magnetic energy which had been stored in the corona through slow stressing applied from the photospheric boundary.  In an idealized model the energy builds up as the coronal magnetic field responds without resistance (every field line line--tied and unbroken).  The flare then occurs as coronal reconnection exchanges those field line footpoints to achieve a lower energy state.  In this process, the footpoints are changed by the reconnection, but the vertical photospheric field in which the field lines are anchored is not.  The potential field from this fixed photospheric field has the minimum magnetic energy possible.  The maximum energy available for release is the amount by which the initial field exceeds this potential field energy, called the {\em free energy}. 
 
 Quantitative simulation of the above scenario has proven to be extremely challenging owing to the vast range of scales involved. Magnetic fields of even modest complexity develop, when stressed, current structures many orders of magnitude thinner than the global length scale \citep{Parker:1972,VanBallegooijen:1985,Longcope:1994}.   A numerical solution therefore requires additional magnetic diffusion to prevent the development of unresolvable current structures.  The corresponding diffusive time is necessarily much shorter than actual diffusive times and generally shorter than the multi-day times governing the stressing phase.  As a consequence, direct simulation, for example by time--dependent MHD solution, includes artificial (diffusive) energy losses competing with the energy build-up.  Few such computations have been capable of demonstrating pre--flare energies comparble to those released by the ensuing flare\citep{Linker:1999}. 
 
 An alternative means of estimating pre--flare energy storage is offered by the Minimum Current Corona model \cite[MCC:][]{Longcope:1996,Longcope:2001}.  This is a quasi-static technique using equilibira, called flux constrained equilibria (FCE), minimizing magnetic energy subject to a set topological constraints composing a subset of all line-tying constraints.  Rather than constraining every pair of footpoints, the MCC groups footpoints into unipolar photopsheric regions and constrains the net coronal flux connecting each pair of regions, called {\em domain fluxes}. Since it uses only a subset of the actual constraints, its fields provide a lower bound on the actual free energy.   As the photospheric regions move relative to one another the potential field above changes as do the fluxes by which it would link region pairs (potential domain fluxes).  Since the actual field is constrained from changing these fluxes, it becomes increasingly different from the potential field and thus gains free energy.  Significantly, this energy is available for release by the violation of topological constraints, a process which can occur on very small spatial scales.
 
 The MCC has been used to estimate pre--flare energy storage in a number of flares \citep{Longcope:1998,Longcope:2007,Kazachenko:2009,Kazachenko:2010, Longcope:2010,Kazachenko:2011}.  A partitioning algorithm was developed to automatically group photospheric flux into distinct regions \citep{Barnes:2005,Longcope:2007b,LBB:2009}.   Provided it is permissible to neglect submergence or emergence of flux, the regions can be taken to move relative to one another but with constant flux.  This simplified scenario was deemed adequate to model pre--flare evolution in several cases  \citep{Longcope:1998,Longcope:2007}, including the landmark Halloween event \citep{Kazachenko:2010}.  When footpoints within a given region move internally, such as during sunspot rotation, additional constraints must be introduced involving the arrangement of footpoints within regions \citep{BeveridgeLongcope:2006,Kazachenko:2009}.  
 
 Flux emergence is a well--known precursor for flares and CMEs \citep{Archontis:2008} and is thus likely to play a role in energy storage for many flares.  It has not yet been accounted for in a complete MCC energy estimate owing cheifly to technical hurdles.  Doing so would require the constraints to be somehow modified to account for flux emergence.  Two prototypical cases were treated by ignoring all but one constraint to which emergence makes an obvious modification \citep{Longcope:2005,Longcope:2010}.  A step toward a more general application was made by \citet{CPLP:2004}, who accounted for emergence and submergence in a computation of coronal reconnection times in the quiet Sun.  Changes in the fluxes of photopsheric regions were used to generate a list of flux changes due to emerging and submerging domains.  The algorithm used had several drawbacks, including a tendancy to assign both emerging and submerging domains to the same photopsheric region. 
 
 The present work introduces a new algorithm by which emerging flux regions may be automatically accommodated in the MCC constraints leading to a pre--flare energy estimate.  Flux changes in  photospheric regions are used to generate a list of domain flux changes due to emergence and submergence, as in \citet{CPLP:2004}.  In this case, however, a single region is linked  to only emerging or submerging domains according to the sense of its own change.  The possibility of artificial photospheric flux changes due to variations in sequential partitioning can be accomodated at the same time using a closely related algorithm.  (This step is independent of the partitioning algorithm.)  We also present a new algorithm for automatically associating domains with separators, a crucial step in the generation of an energy estimate. 
 
 The new methodology is illustrated by applying it to a flare which occured on 16 Oct.\ 2010 and was observed by instruments on the Solar Dynamics Observatory (SDO) spacecraft. New flux emerges into an existing active (NOAA AR 11112) for two days prior to the flare (GOES class M3).  We derive photopsheric magnetic fluxes from a series of 123 line--of--sight magnetograms from the HMI instrument \citep{Scherre:2011,Schou:2011,Wachter:2011} at cadences of $\sim30$ minutes, as discussed in the next section.  In Section \ref{sec:photosphere}, we detail our algorithms for partitioning the set of magnetograms into unipolar regions.  The flux variations in these regions are used to define a set of emerging and submerging domains which are automatically generated by an algorithm explained in Section \ref{sec:change}.  The emerging domains largely resemble those we would have expected based on inspection of the time series, but still contain some physically dubious assignments that must be fixed by hand.  In the future, we would like to completely automate each of these algorithms.
 
 In Section \ref{sec:erg}, we use the magnetograms to derive a coronal topology for the post--flare potential magnetic field and use this to place a bound on the free energy of the pre--flare field.  The post--flare potential field includes a coronal null point whose fan surface encloses one of the newly emerged polarities.  This is a novel feature in the MCC and requires the development of one additional method.  We apply our energy estimate to NOAA AR 11112 in Section \ref{sec:ergapp}, which reveals that most of the pre--flare energy is due to currents passing through the coronal null point. 
 
 \section{\label{sec:partic}Particular Case}
 
\begin{figure}[ht]
   \capstart
   \begin{center}
     \includegraphics[width=\textwidth]{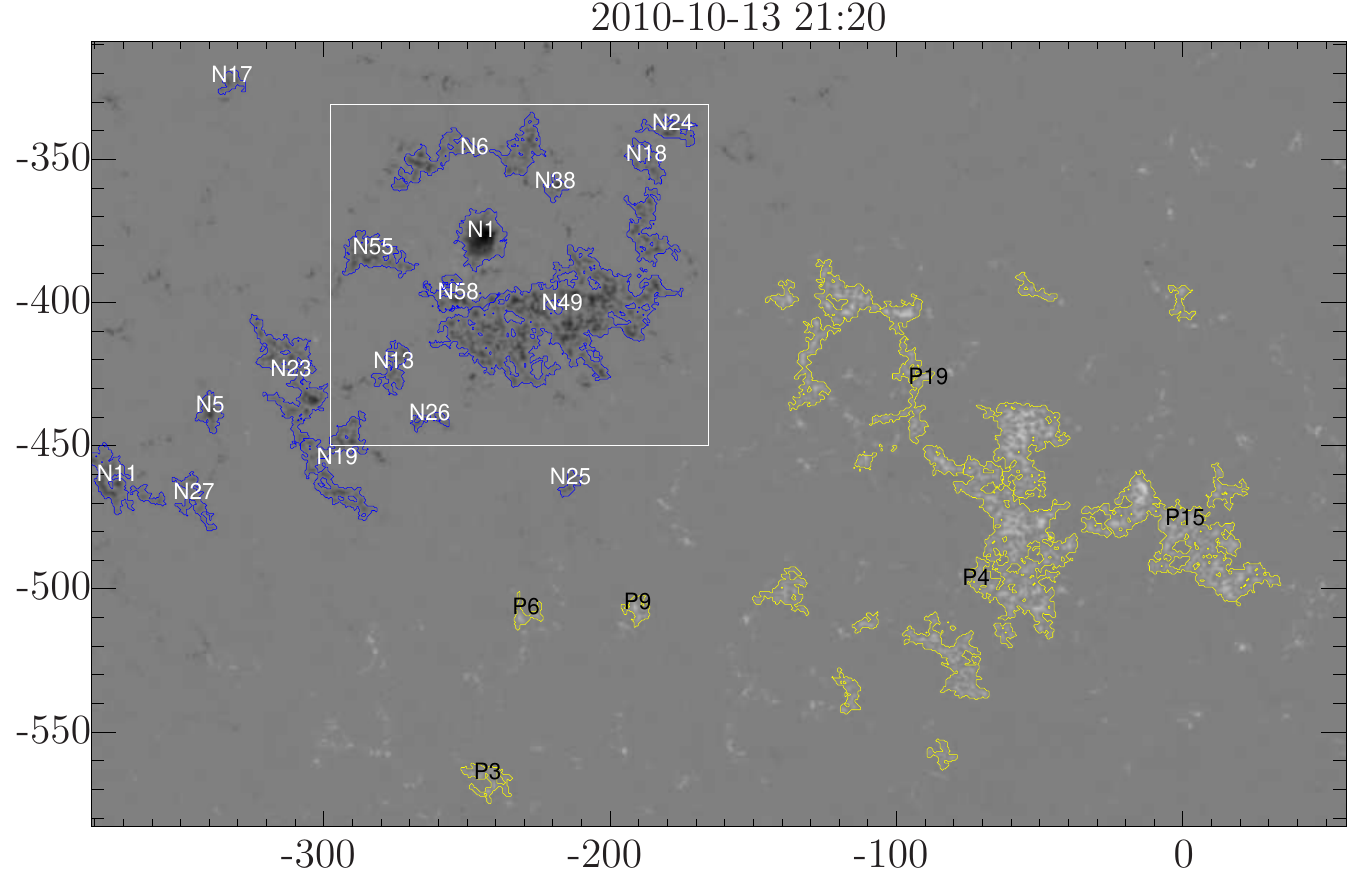}
     \caption[NOAA AR 11112 on 2010-10-13]{\label{fig:oldflux}  HMI LOS magnetogram of NOAA AR 11112 prior to new flux emergence.  The grayscale saturates at $\pm max(\vert B\vert )=\pm 1500.0$G, and the axes are in arcseconds from disk center.  The boxed area encloses the region of flux emergence, which begins $\sim 11$ hours after this magnetogram.}
   \end{center}
 \end{figure}

 Though the methods presented here are of general utility, we apply them, for concreteness, to NOAA Active Region 11112.  We base our calculations on magnetograms taken by the Helioseismic and Magnetic Imager (HMI) on board the SDO spacecraft.  \figref{fig:oldflux} shows an example line--of--sight (LOS) magnetogram from Oct.~13th, prior to flux emergence.  Lighter pixels show positive flux, dark pixels negative flux, and gray pixels zero flux; the grayscale saturates at $\pm 1500.0$G.  The active region originally contains only previously emerged flux as it crosses the Eastern Limb in the Southern hemisphere on October 9th, 2010.  New flux begins emerging around 08:00 UT on the 14th, and is associated with a GOES M2.9 flare several days later, at 19:07 UT on the Oct.~16.  The region shows little activity in the 6 days before the flare.

 Our dataset consists of 123 line--of--sight magnetograms, each consisting of $869\times 544$ pixels with 1.0 arcsecond resolution and approximately 0.5 arcsecond/pixel\citep{Scherre:2011}.  They have a cadence of $\approx$ 2 hours for the first 21 timesteps, from 2010-10-13 00:04 UT to 2010-10-14 18:20 UT, and $\approx$ half hour timesteps thereafter, from 2010-10-14 18:20 UT to 2010-10-16 23:37 UT.  Flux emergence begins around timestep 21.  As shown by the supplementary media, a movie of the LOS magnetograms, the photospheric changes are well characterized by these two cadences.  Our data window has flux imbalance $< 10\%$ over the three days leading up to the October $16^{\hbox{\scriptsize th}}$ flare.

\begin{figure}[ht]
   \capstart
   \begin{center}
     \includegraphics[width=0.7\textwidth]{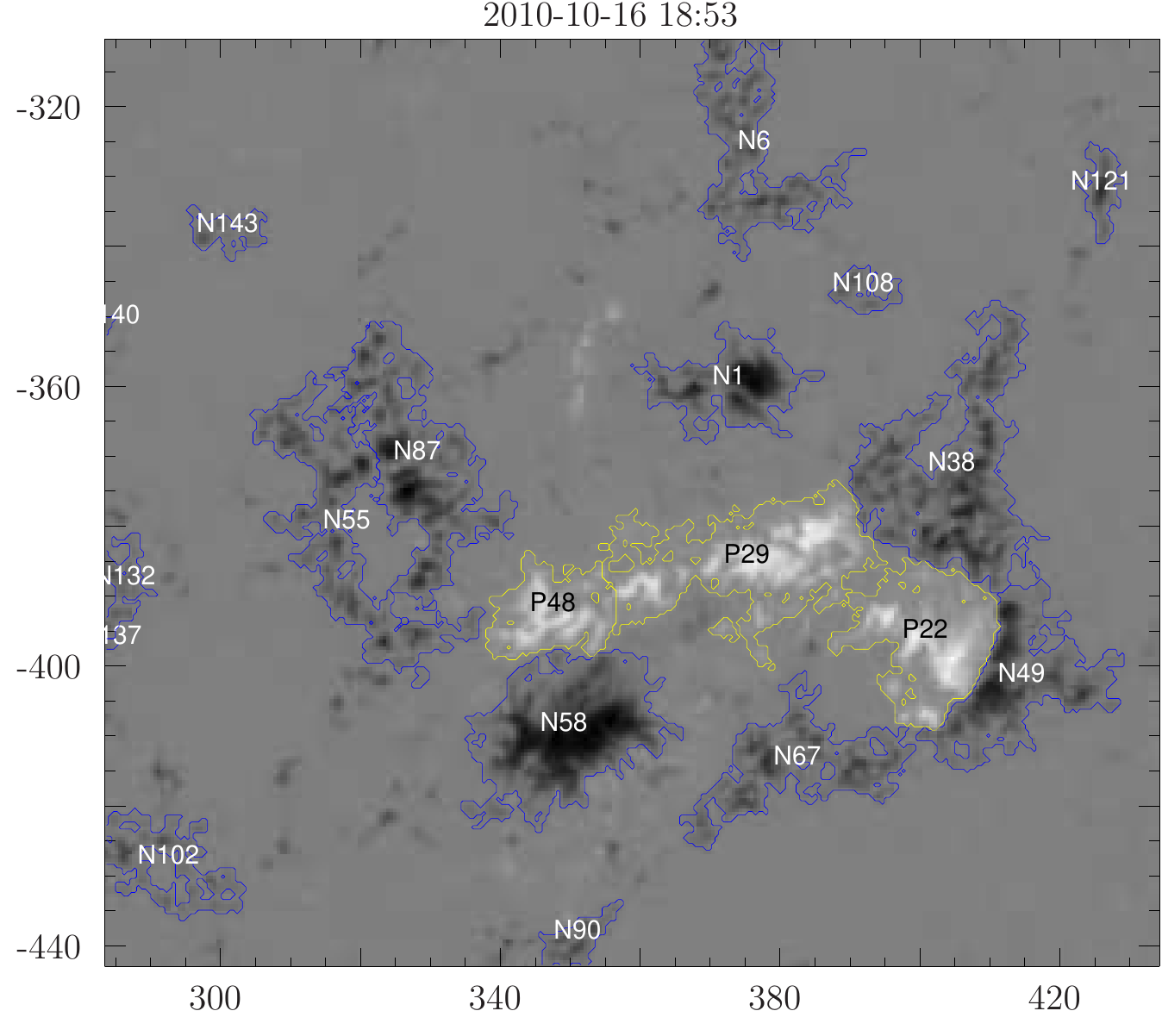}
     \caption[Newly emerged flux within the ring]{\label{fig:newflux}  A zoomed view of LOS magnetogram of the emerging flux region, boxed in \figref{fig:oldflux}, on Oct 16th, 2010, 18:53UT, $\sim$ 10 minutes prior to the M2.9 flare.  The grayscale saturates at $\pm -1176.10$G, and axes are in arcseconds from disk center.}
   \end{center}
 \end{figure}

 The old flux region has a polarity inversion line, running from Southeast to Northwest, which coincides with a filament as seen in data from the Global High--Resolution H$\alpha$ Network, as well as SDO EUV images at 94, 131, 171, 193, 211, 304, and 335 \AA.  This filament appears unaffected by the 19:07 UT flare.  To the North of the filament, the diffuse, old--flux, negative polarity field includes a curious ring of flux surrounding a strong core, boxed in \figref{fig:oldflux}.  The new flux of both polarities emerges completely within this ring, sweeping the old negative flux out of its way as it carves out a space for itself.  This results in the (zoomed in) field shown in \figref{fig:newflux}, roughly 10 minutes before the flare.

 As the new flux emerges it snow--plows the old flux in front of it, the old flux concentrates and a strong horizontal gradient in the LOS field develops between old and new flux, as seen near S 400'', W 410'' in \figref{fig:newflux}.   The flare is centered on this strong polarity inversion line.  Post--flare loops connect the newly emerged negative flux (N55, N58, N87) within the ring to the diffuse, positive flux to the West (P4, P15, P19).  We therefore believe that any analysis of this flare's energetics must explicitly account for flux emergence.

 Our analysis naturally splits into two parts.  The first parts works directly with the photospheric magnetic field measurements.  We first partition the magnetic field at all times into unique, unipolar regions.  Second, we track these regions and enforce a consistent labeling scheme as they change geometry over time.  Geometric changes include changes in size, shape, orientation, and total flux; merging and splitting of regions; and the complete emergence or submergence of regions.  The final step quantifies flux emergence and submergence between pairs of photospheric sources using a novel algorithm.

 In the second part of our analysis, we use our characterization of the photospheric field as a set of unique unipolar sources to describe the topology of the coronal field.  Each distinct photospheric region is replaced by a point source of the same flux located at the region's flux--weighted centroid.  We then find the flux within each potential field domain, and the null points and separators of the potential field.  Changes in the potential field's domain flux relative to the actual domain fluxes, assumed fixed, indicates the storage of magnetic energy in excess of the potential field.  In an energy--minimized field, the free energy stored by the field is manifest by current ribbons that develop along separator field lines.  For brevity, we will use the phrase ``energy stored in the separator'' as a shorthand for ``free energy stored in the magnetic field which manifests itself as a a current ribbon at the location of a separator.''  The amount of current flowing along these ribbons places a bound on the free magnetic energy of the system.

 \section{\label{sec:photosphere}Characterizing the photospheric field}
 
 We begin work on the photospheric data by partitioning the magnetogram using the gradient--based tessellation algorithm described by \citet{Barnes:2005}.  The initial LOS magnetograms are converted to a vertical field by assuming a radial field at each pixel, which amounts to dividing by $\cos(\theta)$, where $\theta$ is the polar angle from disc center.  When determining the flux in each region, we must also account for foreshortening within each pixel, dividing by a second factor of $\cos(\theta)$.  We then convolve all vertical field, $B_z=B_{LOS}/\cos^2(\theta)$, with the Green's function for a potential extrapolation up to a height $h$ from an unbounded plane:
 \begin{equation}
   K_h(x,y) = \Frac{h/2\pi}{(x^2+y^2+h^2)^{3/2}}.
 \end{equation}
 To reduce the effects of noise, we neglect all the convolved field below a threshold of $\vert B_{th}\vert = 50$G.  Using the smoothed field, we assign a unique label to all local maxima and every pixel strictly downhill with respect to $\vert K_h\star B_z\vert$ from each maxima.  Internal boundaries in unipolar regions are eliminated---multiple regions are merged into a single region---when the saddle point $\vert B_z\vert > \hbox{min}(\vert B_{pk}\vert)-B_{sad}$. $B_{pk}$ is the greatest vertical field strength of the saddle's surrounding peaks, and $B_{sad}$ is a threshold value.  At the end of this process, each data pixel has an associated label.  We call the set of all labels at a given time a mask, and the set of masks at all times a mask array.

 Our goal in this section is to make the mask array consistent from timestep to timestep.  In particular, we wish to determine how the newly emerged flux interacts with the old flux, and how that process affects the energetics of the active region.  Our primary concern is therefore to distinguish between new and old flux, which is a marked departure from previous, similar investigations\citep{Kazachenko:2009,Kazachenko:2010,Longcope:2007b}.

\figref{fig:newflux} shows an example where the boundary of each unique mask region and its label (P1, N1, P2, \ldots) have been plotted over the line of sight magnetogram data.  All pixels and regions containing flux below our thresholds are given a label of 0.  This figure covers the ring feature of old flux---concentrated in regions N1, N6, N38, N49, N55, and N67---and the newly emerged flux in regions P22, P29, P48, N58 and N87.  The supplimentary material shows that these definitions are not necessarily strict.  For instance, N58, N67, and N87 initially all break off of N49.  Both N58 and N87 are quickly dominated by newly emerged flux, while N67 remains primarily distinct.  We refer to ``new flux'' or ``old flux'' regions based on the dominant type of flux determined by inspection of the timeseries (see supporting media).

 For the present analysis, we have set $h=1.0$Mm, $\vert B_{th}\vert =50G$, and $B_{sad} = 0.8\times B_{pk}$.  We also established a minimum flux of $2.6\times 10^{19}$Mx $(20000$ Gauss px\tothe{2}, after accounting for pixel foreshortening) for a solitary region.  Our field of view contains unsigned flux of order $10^{22}$Mx above the 50 Gauss threshold, so, a region must contain at least 0.2\% of the total unsigned flux before inclusion in our algorithms.  For a discussion of how the parameters $h$, $B_{th}$, and $B_{sad}$ effect the final partitioning, see \cite{LBB:2009}.
 
 After partitioning each magnetogram, we attempt to associate the partitions at one timestep to those in the next.  To begin with, each region is characterized by its net signed flux and centroid location:
 \begin{equation}
   \label{eq:psi}
   \psi = \int_\mathcal{R} B_z(x,y)\, dx\, dy \qquad \bar{\vect{x}} = \psi^{-1}\int_\mathcal{R} \vect{x} B_z(x,y)\, dx\, dy
 \end{equation}
 For short, we call a region's total flux at the flux--weighted centroid the region's associated \emph{pole}.
 
 As a first pass at creating consistently identified regions, we calculate the distance between each pair of centroids at two consecutive timesteps.  If Centroid A at time $i$, $\vect{x}_A^i$, is closest to Centroid B at time $i+1$, $\vect{x}_B^{i+1}$, and B has A as its closest neighbor, and that distance is less than a threshold (10Mm), then we conclude that Region B is Region A.  

Both this simple method of association, and exploration of the partitioning parameter space, result in mask arrays that are not to the quality required for MCT analysis of an emerging flux region.  We therefore developed two more sophisticated automatic procedures.  \fnc{rmv\_flick} and \fnc{rmv\_vanish}, described below, provide refinements to the minimum--distance identifications.  Both were heuristically developed to address a prevalent type of inconsistent region labeling.  The first algorithm deals with a region that exists for just a single timestep, but is clearly part of another region.  The second deals with regions that change labels from one timestep to the next.  These problems arise, for instance, when the choice of parameters poorly represents a subset of the timeseries, or when two regions merge or split, and their centroid locations between two timesteps are very different.
 
 For clarity, our examples in the following subsections do not directly use masks from our time series, and are only meant to illustrate the action of each algorithm.  Both algorithms operate directly on pixels in the mask structure, so that there are no physical scales involved.  The algorithms only relabel nonzero elements of the mask, so that no external mask boundaries are modified.

 \subsection{\label{sec:rmvf}\fnc{rmv\_flick}}
 
 \begin{figure}[ht]
   \capstart
   \begin{centering}
     \includegraphics[width=0.5\textwidth]{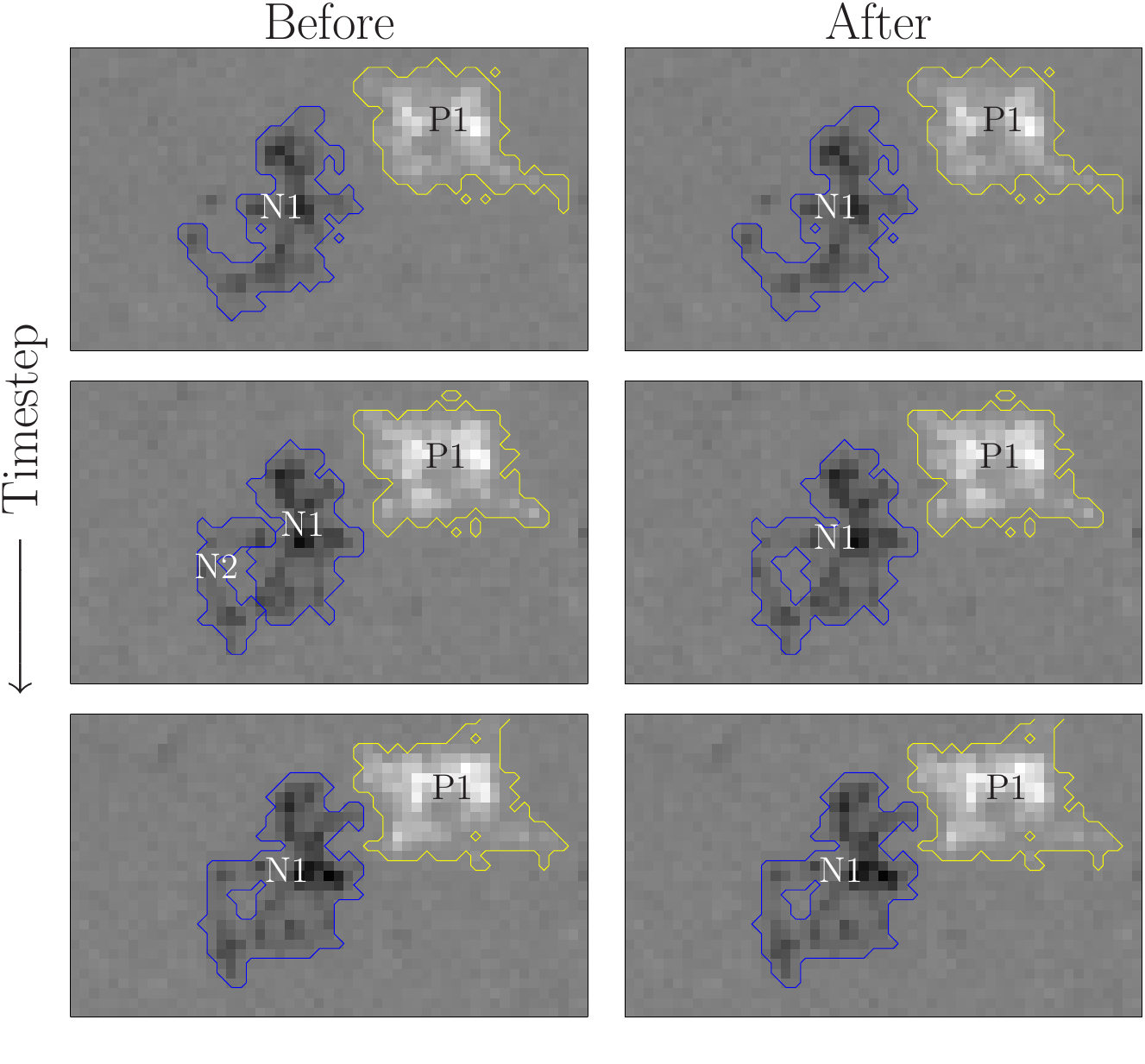}
     \caption[Example of \fnc{rmv\_flick}]{\label{fig:rmvf}The first column demonstrates the problem of ``flickering.''  The middle timestep contains a new created region, $N2$, inside of the area controlled by region $N1$ in the surrounding timesteps.  $N2$ only exists for that single timestep, its area reverting back to $N1$ in the third timestep.  The right column shows the effect of running \fnc{rmv\_flick} on this region: in the middle timestep, $N1$ has been completely restored in place of $N2$.}
   \end{centering}
 \end{figure}
 
 \fnc{rmv\_flick} attempts to smooth the temporal structure of our mask arrays by better associating regions that last for just a single timestep.  This happens when the saddle points in the tessellation algorithm slosh back and forth over the threshold value.  \figref{fig:rmvf} shows an example of this problem.  Region N2 pops into existence in the middle timestep, occupying some of the territory ascribed to region N1 in the first and third timestep.  We see that this is not a ``real'' new region, but should instead be considered part of N1 for the entire time.  \figref{fig:rmvf} column two shows the result of running \fnc{rmv\_flick} on this data.
 
 The \fnc{rmv\_flick} algorithm works on a sliding, three timestep window (initial, middle, and final timestep), and in two steps.  In the first step, we find all regions $\Theta_i$ that have nonzero flux only in the middle timestep, and find all mask pixels in the initial and final timestep which overlap $\Theta_i$.  We take each overlapping region $\Theta_A$ and relabel those pixels of $\Theta_i$ which are overlapped by $\Theta_A$ in the initial and final timesteps as $\Theta_A$.

 Because our source regions change shape and size (pixel count) over time, we will usually have leftover pixels after performing the above bulk relabeling.  While leftover pixels remain, we relabel the pixel of $\Theta_i$ that currently borders the greatest number of pixels of a single neighbor to that neighbor's label.  We have found that, when multiple regions overlap a flickering region, this method divides those pixels between the overlapping regions in decent proportion to their past and future ``control'' of flickering territory.

 \subsection{\label{sec:rmvv}\fnc{rmv\_vanish}}

 \begin{figure}[ht]
   \capstart
   \begin{centering}
     \includegraphics[width=0.5\textwidth]{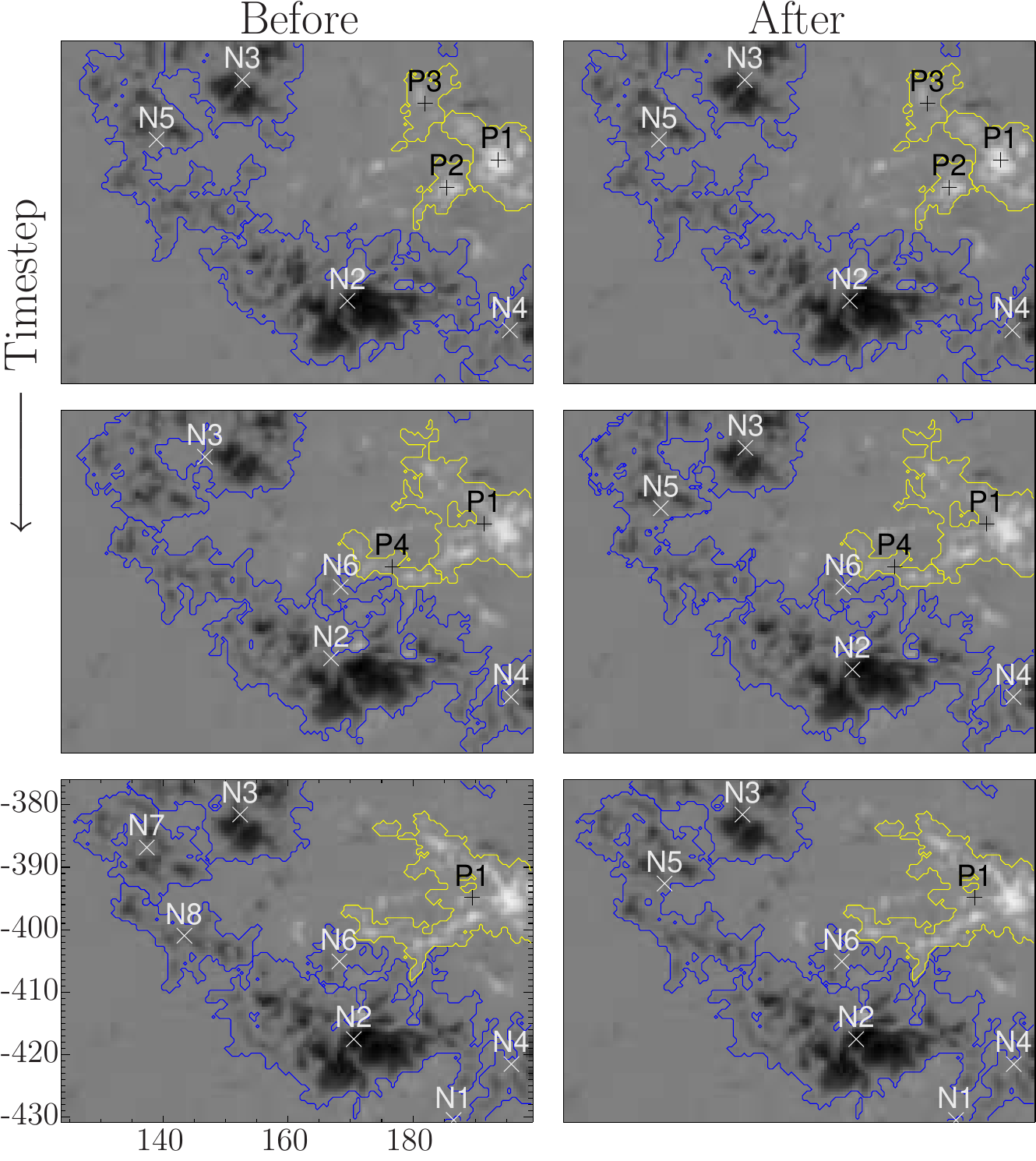}
     \caption[Example of \fnc{rmv\_vanish}]{\label{fig:rmvv}The left column demonstrates the problem of ``vanishing,'' in which a single region---N5 in the top left panel---cycles through series of different labels.  The right column shows the effect of running \fnc{rmv\_vanish} on this data.  The $x$ and $y$ axes are in arcseconds from disk center, though the physical scale is irrelevant to the algorithm's operation.}
   \end{centering}
 \end{figure}
 
 With \fnc{rmv\_flick} we dealt with regions that existed for just a single timestep.  We now deal with the conceptual inverse problem, where a region exists for some time and then suddenly changes names in the next timestep.  \figref{fig:rmvv} demonstrates the problem, as the region labeled N5 in the top left panel (and previous, undepicted times) cycles through a series of names and partitions within a few timesteps.
 
 We again begin by considering the mask array in a sliding, three timestep window.  We find all non--zero flux regions within the window and determine which regions exist in which timesteps.  For every region that disappears in the middle timestep, we project its area from the first timestep into the middle and last timesteps.  We then find all new regions, in both the middle and final times, that overlap with that projected area.  Finally, we relabel any new region whose centroid lays within this projected area.  \figref{fig:rmvv}, right column shows the effect of running \fnc{rmv\_vanish} on data.
 
 \subsection{\label{sec:mskresults}Results of the consistency algorithms}
 
 \fnc{rmv\_flick} and \fnc{rmv\_vanish} each use a sliding, three timestep window, so that it takes many repetitions of each for updated information propagate from the beginning to the end of the mask array.  We must therefore repeat each of the algorithms many times.  At the same time, \fnc{rmv\_flick} acts upon labels that last for just a single timestep; eventually, all such problems will be fixed.  Taking these processes into account, we find that the mask arrays best approximate the photospheric field evolution when we switch back and forth between the two algorithms at the beginning of the process, and run just \fnc{rmv\_vanish} many times at the end of the process.  For our NOAA AR 11112 data, we ran \fnc{rmv\_flick} a total of 35 repetitions, and \fnc{rmv\_vanish} a total of 1058 repetitions.
 
 Together, these two algorithms accomplish about nine--tenths of the work in creating a consistent set of tessellated masks.  Part of the remaining tenth may be accounted for by a boundary--shift algorithm, described below, between abutting regions of the same polarity.  Even this does not accurately represent to data, and the final changes to the mask array required for a consistent time series must be done by hand.  These changes again mostly involve the placement of the boundary between abutting regions of the same polarity, caused by a failure to distinguish between old flux and actively emerging flux.  These boundaries are adjusted manually until each region's flux evolves smoothly.  

\begin{figure}[ht]
   \capstart
   \begin{centering}
     \includegraphics[width=0.8\textwidth]{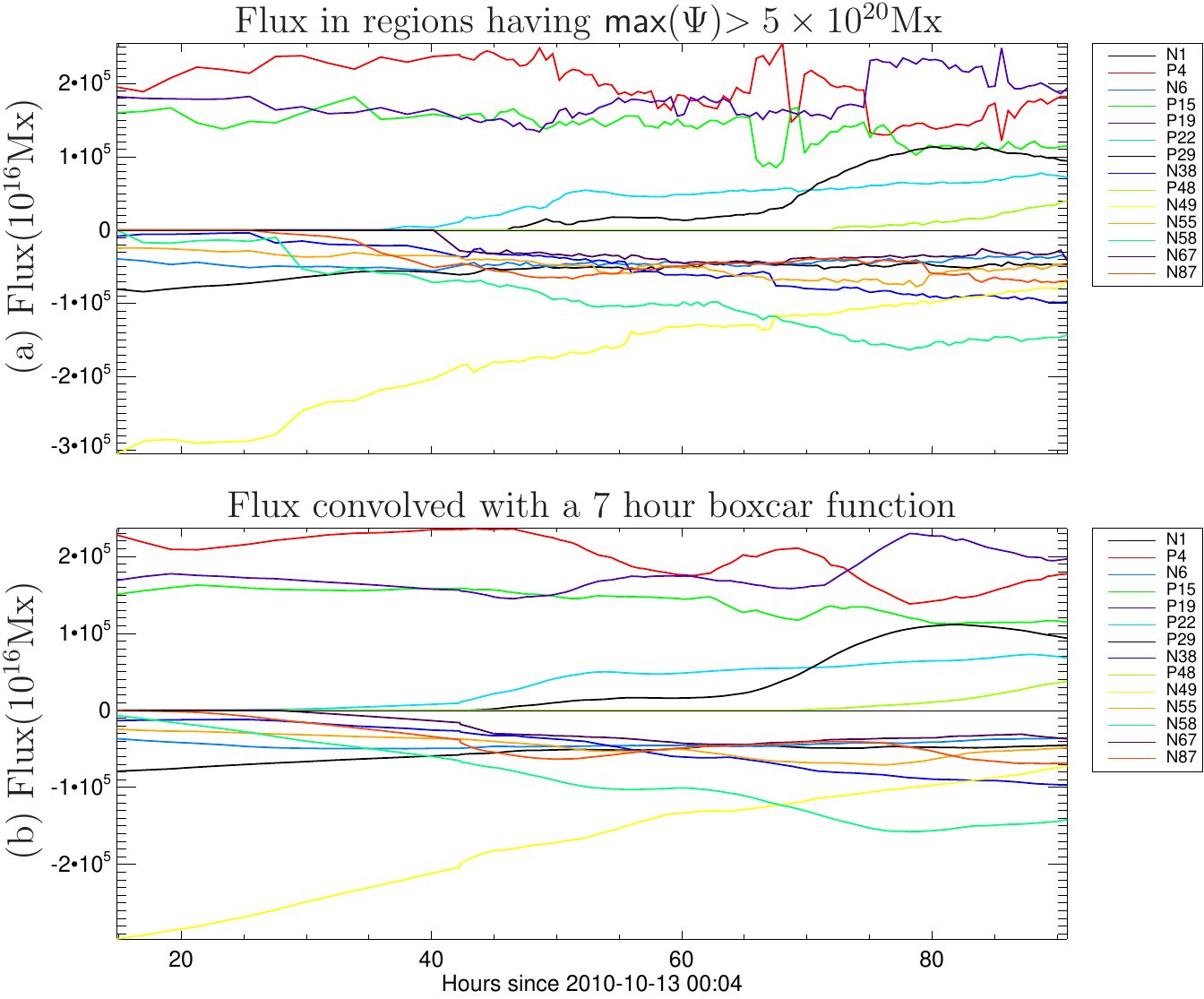}
     \caption[Flux within each region of NOAA AR 11112]{\label{fig:fluxplot} (a) Flux in regions with more than $5\times 10^{20}$Mx.  Large, opposing spikes between pairs of curves, for instance between P15 and P19, show boundary shifts between adjacent regions.  (b) Result of smoothing the data by convolution with a 7 hour boxcar function.}
   \end{centering}
 \end{figure}

 \figref{fig:fluxplot} top shows the vertical flux in each (high--flux) region.  This includes editing of the masks by both the automatic algorithms and by hand.  Note that, even after a set of consistent masks have been created, the flux within each region is still rather noisy.  To counteract this, we smooth our data with a 7 hour (13 timestep) boxcar function.  The result is shown in \figref{fig:fluxplot} bottom.
 
 \section{\label{sec:change}Change--in--Connectivity Algorithm}
 At this point, we transition from a mostly qualitative treatment of the photospheric field's geometry to a quantitative, topological analysis of a model coronal field: the Minimum Current Corona (MCC) Model developed in \citet{Longcope:1996, Longcope:2001}.  We begin by replacing every region in the photospheric field with a magnetic pole defined by the region's total flux and centroid location, as in \eqref{eq:psi}.  With one exception, the rest of this work deals only with the poles and connections between them.  

 The total flux in each pole must be connected to some number of other poles of opposite polarity.  This defines the system's connectivity: the undirected, weighted graph, where each vertex of the graph is a pole, and the weight of each edge defines the connectivity between the two vertices.  It happens that the graph of NOAA AR 11112 is \emph{simple} on the eve of the M3 flare, so that each pair of vertices has no more than one associated edge\footnote{This is not generally true for solar magnetic topologies, where two poles may have multiple topologically distinct edges divided by redundant separators \citep{BeveridgeLongcope:2005,Parnell:2007}.  At every timestep throughout the time series for which we calculated AR11112's connectivity, we found it to be simply connected.  The next section will define these topological terms in more detail.}.  We refer to edges of the graph as domains.  We may additionally define a connectivity graph for the flux \emph{change} between two timesteps: if a pole's flux increases, that increase must be distributed among its domains with opposite polarity poles.  This is the topological entity we now quantify.

 The flux in a single photospheric region, and hence that region's associated pole, may vary in two ways.  The change is either a true evolution of the field by submergence or emergence through the photosphere, or it is a transfer of flux to/from another like--signed region via a shift in the boundary between the regions.  That this latter type still exists in our ``consistent'' data can be seen in regions P4, P15, and P19 in the smoothed data of \figref{fig:fluxplot}.  These three regions compose the large, diffuse positive region to Solar West, as seen in \figref{fig:oldflux}.  

\begin{figure}[ht]
   \capstart
   \begin{centering}
     \includegraphics[width=\textwidth]{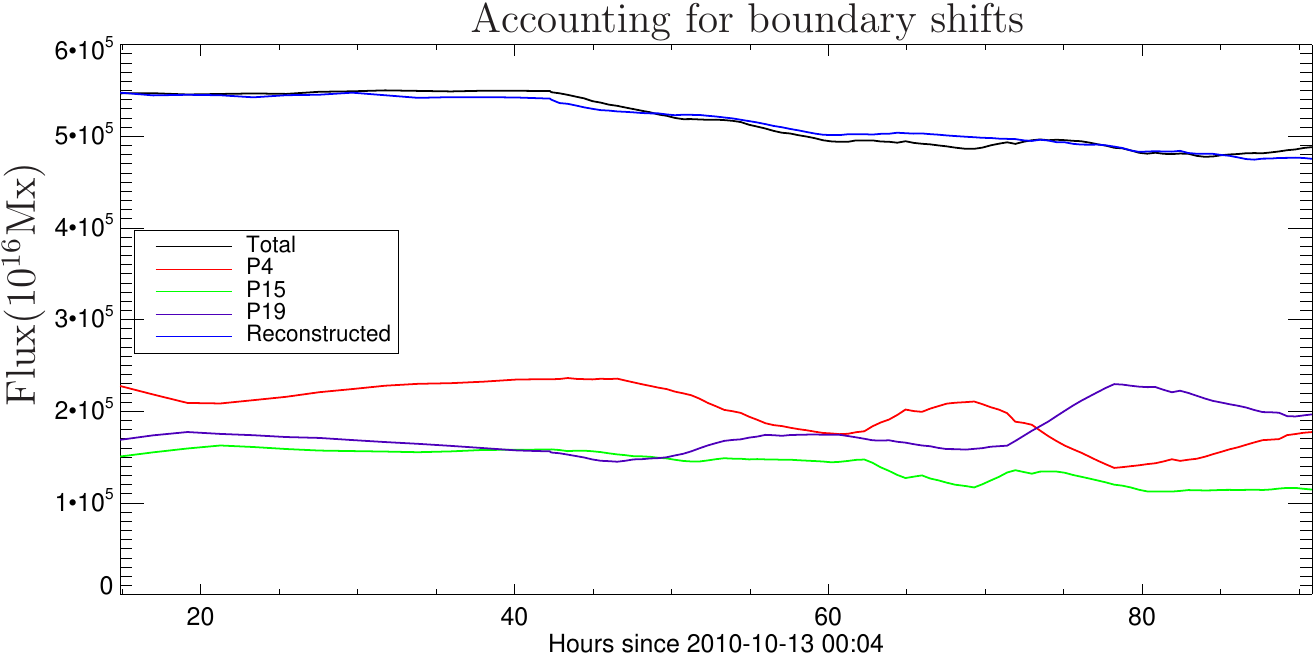}
     \caption[Result of \fnc{bndry\_shift}]{\label{fig:bndry}  The submergence of regions P4, P15, and P19 (red, green, purple), shown in the total flux of the regions (black).  Blue shows the calculated submergence by removing flux change due to boundary shifts from the total flux change.}
   \end{centering}
 \end{figure}

 If we focus on each of these regions individually, the red, green, and purple lines in \figref{fig:bndry}, we see several rapid changes in the flux of each.  However, if we look at the total flux in these three regions---the black line in \figref{fig:bndry}---it steadily decreases: the rapid increases and decreases in \figref{fig:fluxplot}b are almost solely due to shifting boundaries between the three.  Further, the steady decrease shown in the black line is matched by decreases in the old--flux part of the flux ring, regions N1, N6, N38, N49, N55, and N67 in \figref{fig:newflux}: the old positive flux is ``submerging'' with the old negative flux.  The blue line is a demonstration of the effectiveness of the forthcoming algorithms, and we will revisit it shortly.

\subsection{Quantifying flux change due to boundary shifts}

 In order to isolate the different varieties of flux change, we have developed an algorithm which estimates the change in each region's flux due only to boundary displacements between adjacent, like--signed regions.  These changes necessarily come in pairs: what one region loses in a boundary shift, another gains.
 
 Our algorithm works as follows.  Consider a set of like--signed poles $\{P\}$, and split $\{P\}$ into submerging and emerging sets, $\{P_\downarrow\}$ and $\{P_\uparrow\}$ respectively.  Each pole has an associated flux change, $\Delta^i\psi\equiv \psi^{i+1}-\psi^i$, with $\psi^i$ the flux at time $i$ given by \eqref{eq:psi}.  We iteratively find the pole with smallest unsigned flux change, $P_s$, and the pole with closest centroid $\mathbf{x}$ and opposite sense flux change, $P_c$.  The flux change between these two poles is canceled, $\Delta\psi_c\rightarrow \Delta\psi_c-\Delta\psi_s$ and $\Delta\psi_s \rightarrow 0$, and the cancellation is recorded in a change--in--connectivity matrix: $\Delta^i\Matrix{M}_{c,s}=-\Delta^i\Matrix{M}_{s,c} = \Delta\psi_s$.  This process is repeated until no more connections can be made.  

 $\Delta^i\Matrix{M}$ is antisymmetric, and the $\{j,k\}$\tothe{th} element is the flux--change of the edge joining $j$ and $k$.  Each row $j$ describes the flux change for a given pole, and each column $k$ in a row records how much of pole $j$'s total flux change is in partnership with $k$.  If all of pole $j$'s flux change is due to boundary shifts, then $j$'s total flux change is given by summation along the $j$\tothe{th} row of $\Delta^i\Matrix{M}$:
 \begin{equation}
   \psi^{i+1}_j = \psi^i_j + \sum_{k}\Delta^i\Matrix{M}_{j,k}.
 \end{equation}
 In this way, we have paired as much shrinking flux with like--signed increasing flux as possible, but have only used the total flux and centroid location of each region.
 
 We now refer back to the photospheric mask for the last time.  For each region in the mask array we find every like--signed region with which it shares a border: call this subset of poles $\{b\}$.  To allow for regions separated by a few pixels to share a boundary\footnote{cf. \figref{fig:newflux} and supporting media: at various timesteps, N38 and N49 are separated by several pixels of low field, yet there is a clear transfer of flux across their shared boundary.}, we pad each region by 5 pixels and look for overlap with other regions.  The value of $\Delta^i\Matrix{M}$ for each pair of such regions is then the amount of flux they exchanged through a shift in the mask boundary.  
 
 We have chosen to distribute the flux--change and then restict to the subset $\{b\}$, rather than the reverse, in order to avoid the ill--defined problem of assigning shifts when one region shares boundaries with multiple other regions, which in turn share boundaries with multiple other regions, and so on.  We see this, for instance, in the triple boundary between regions P4, P15, and P19.  Qualitatively, any discrepancy is small given that we connect regions in a nearest--to--farthest order, and nearby regions tend to share boundaries.

 The sum of all such changes for a given pole gives that pole's total flux change due to boundary shifts.  All other change must be ascribed to submergence or emergence through the photosphere.  We may describe this mathematically as
 \begin{equation}
   \psi^{i+1}_j = \psi^i_j + \sum_{b}\Delta^i\Matrix{M}_{j,b} + \sum_k\Delta^i\Matrix{S}_{j,k}.
 \end{equation}
 The middle term on the RHS is the pole's boundary change, determined in this section; the final term quantifies the pole's flux change due to submergence and emergence through the photosphere, which we now quantify.

 \subsection{\label{sec:qsef}Quantifying flux change due to submergence and emergence}

 Our algorithm for determining the flux change due to submergence and emergence rests on two assumptions.  First, pairs of poles submerge and emerge together, and second, a single region can either consist of submerging or emerging flux, but not both.  These assumptions naturally break our connectivity--change graph into two disconnected subgraphs: one consisting of submerging vertices, the other emerging.  Each subgraph is composed of both positive and negative polarity poles, as flux connects only regions of opposite polarity.  

 In determining the flux change in each domain, we use the same algorithm as for boundary shifts, only with the roles of polarity and sense of change reversed: these domains connect opposite polarity poles with same sense flux change, and represent pairs of poles submerging or emerging together.  The graph generated in this manner reflects our physical intuition about how these systems should interact.  Namely, we expect poles to be more connected to those poles closest to them, rather than those further away.   We also expect that poles with little flux change should not have that small change spread between a large number of partners.
 
 \begin{figure}[ht]
   \capstart
   \centering
   \includegraphics[width=\textwidth]{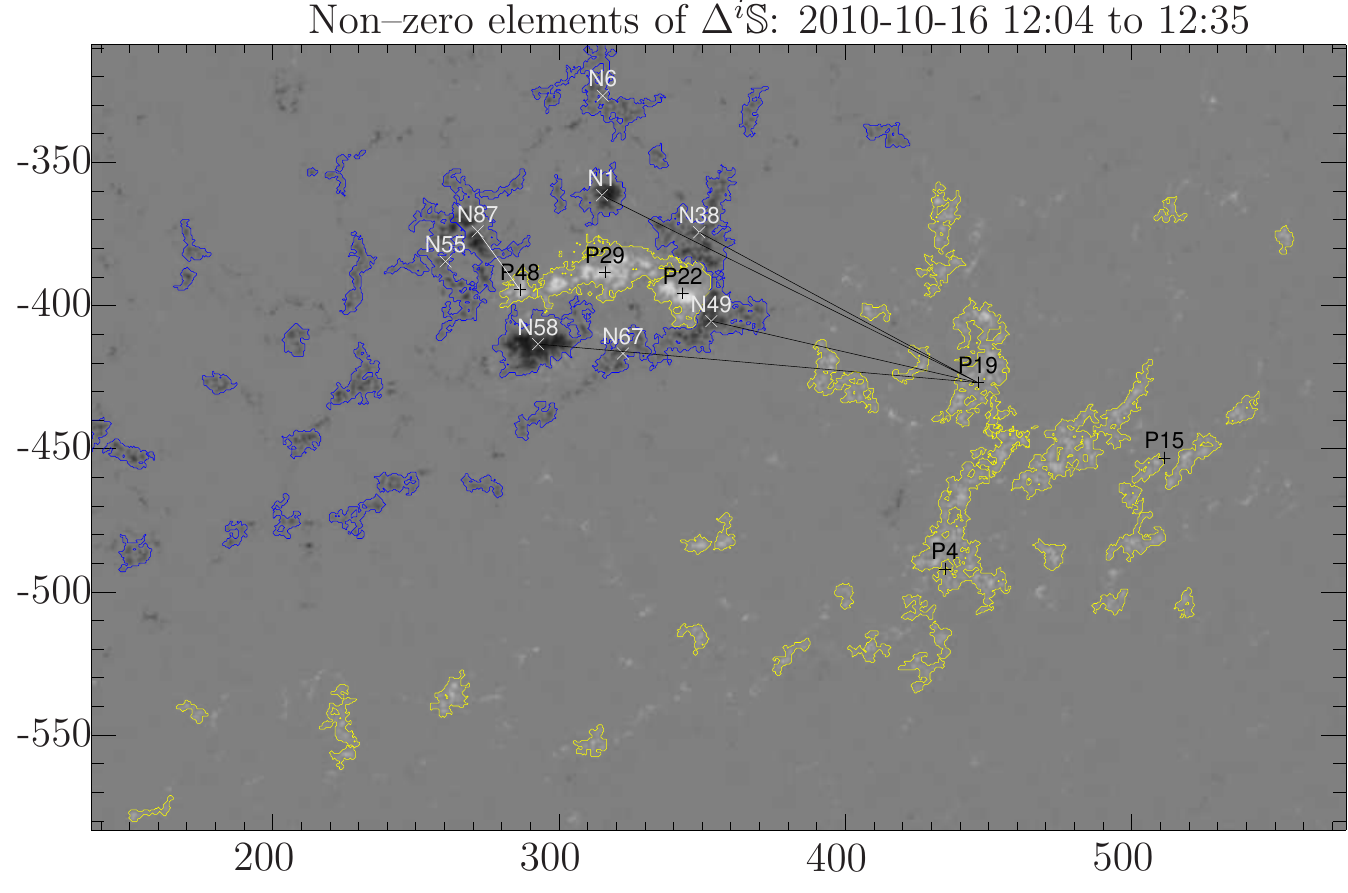}
   \caption[Progression of the connectivity--change algorithm]{\label{fig:smat} Example map of sub/emerging flux, as generated by the algorithm of \S \ref{sec:qsef}.  Black lines show submerging regions, white lines emerging.  Labels mark the centroid locations of the high--flux regions of \figref{fig:fluxplot}, which also contribute to the topological analysis of \S \ref{sec:ergapp}.}
 \end{figure}
 
 We begin by finding the flux difference for each pole between times $i$ and $i+1$, from which we subtract the change due to boundary shifts, found previously.  The remainder is each pole's flux change budget, which must be paired with other poles.  To accomplish this, we again iteratively find the least changing pole and cancel its flux change with its closest appropriately signed and sensed neighbor.  This cancellation is the weight of each edge and is recorded in the source--change matrix, $\Delta^i\Matrix{S}$.  Just like the boundary shift matrix, $\Delta^i\Matrix{S}$ is defined such that, if all of a pole $P_j$'s flux--change between times $i$ and $i+1$ is due to real photospheric changes, then $\psi_j^{i+1} =\psi_j^i + \sum_k\Delta^i\Matrix{S}_{j,k}$; that is, summation over all columns of a given row returns the total flux--change for that row's pole.

 While the combination of our algorithms dealing with boundary shifts and emergence do quite well in most circumstances, occasionally they produce unphysical connections.  Consider the timestep shown in \figref{fig:smat}.  Here, the emerging positive regions (P22, P29, P48) each have some flux change, and our knowledge of the location of emergence says that this change is most likely a combination of boundary shifts and emergence for P48 and P29, as both N87 and N58 are also actively emerging, and primarily boundary shifts for P22.  However, if the boundary shift between P22 and P29 is larger than the emergence of P48, then our boundary algorithm ascribes the boundary shift to P29--P48, leaving the emergence mostly between P22 and N87, which is clearly unphysical.  For this reason, we allow for ``preferred connections'', which accounts for emergence between a subset of poles first, and then proceeds to calculate boundary shifts and submergence and emergence for all poles.  For this region, we first pair emergence between regions P22, N55, N58, and N87 for the first 3 days (77 timesteps), and then between regions P48, N55, N58, and N87 during the remaining time.  As with the generation of consistent mask structures in the previous section, we ultimately wish to automate this entire process.  However, until we develop algorithms that accurately capture the physical evolution of emerging regions, we will rely on a combination of automatic proceedures and ad hoc prescriptions to most accurately model these systems.

 \figref{fig:smat} shows a graphical representation of the changing domain fluxes due to submergence and emergence between 12:04 and 12:35 UT on Oct 16th, 7 hours before the flare.  Black lines show submerging domains, white lines emerging.  Table \ref{tab:smat} shows the flux change in each domain.  Regions P4, P15, P22, P29, and N55 have flux change (for these timesteps) due solely to boundary displacements with neighboring regions.

 As discussed above, we begin pairing P48, which increased by $1506.4\times 10^{16}$Mx.  The closest (and only) negative pole with the correct sense of flux change is N87, which increased by $849.9\times 10^{16}$Mx.  These two poles are paired, the domain flux--change is set to $849.9\times 10^{16}$Mx, and P48's flux budget is reduced to $565.5 \times 10^{16}$Mx.  There are no more available connections between our preferred poles, so we next account for boundary shifts.  After doing so, the least changing region is N38, which loses $30.6\times 10^{16}$Mx.  The closest positive pole with the correct sense of flux change is P19, which lost $1528.9\times 10^{16}$Mx.  These two poles are paired, the domain flux--change is set to $30.6\times 10^{16}$Mx, and P19's flux budget is reduced to $1498.3\times 10^{16}$Mx.  We again select the least changing region (N49), and the process continues until no more pairings can be made.  The final results for this timestep are shown in Table \ref{tab:smat}.

 \begin{table}[ht]
   \centering
   \caption{\label{tab:smat}Non--zero changes in domain fluxes between 12:04 and 12:35 on 2010-10-16}
   \begin{tabular}{lcr}
     \tableline\tableline
     Domain & Sense &  $\Delta^{12:04}\Matrix{S} (10^{16}$Mx) \\
     \tableline 
     P19---N1 &  submerge &     245.764   \\
     P19---N38 &  submerge &     30.6279 \\
     P19---N49 &  submerge &     145.569 \\
     P19---N58 &  submerge &     257.030 \\
     P48---N87 &  emerge   &     849.874 \\
     \tableline
   \end{tabular} 
 \end{table}

 The pairings represent a compromise between reasonable inference and mathematical necessity.  Due to their proximity, and common increase, it is reasonable to assume P48--N87 are feet of the same emerging flux tubes.  P19, on the other hand, is relatively isolated but has steadily decreasing flux.  It is necessary to pair it with decreasing negative poles, which turn out to be arranged along the periphery of the old polarity.  Being separated by $\sim100''$, these regions cannot be ``submerging'' in any real sense.  However, we believe this decrease represents steady cancellation with the small scale field surrounding these old--flux regions.  This process should be studied in detail in another investigation, but for the moment we note that it is well represented by a reduction in, for instance, the P19--N49 domain flux.
 
\begin{figure}[ht]
   \capstart
   \centering
   \includegraphics[width=\textwidth]{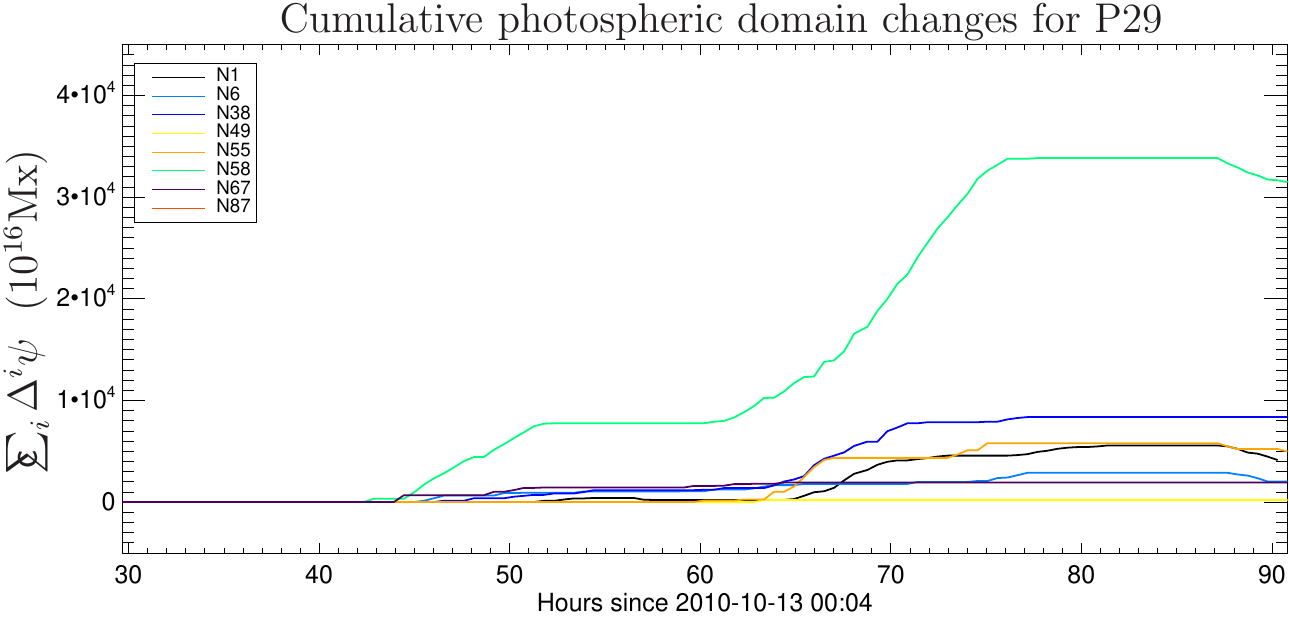}
   \caption[Cumulative changes in P29's domain fluxes]{\label{fig:p29dom}Cumulative changes in each of P29's domain fluxes, as calculated with the algorithm of \S \ref{sec:qsef}.  We find that P29 primarily emerges with N58 and N38.}
 \end{figure}

 We repeat this process for each pair of consecutive timesteps.  \figref{fig:p29dom} provides a representative example, showing the cumulative domain flux changes (designated by the symbol $\csum$) due to the emergence, and slight amount of submergence near the end of our analysis, of P29, one element of the newly emerging flux.  We see here that it emerges primarily with region N58, part of the newly emerging flux.

 The effectiveness of the combined algorithms can be appreciated in \figref{fig:bndry}, showing the three componenets of the diffuse Western region.  The blue line illustrates the effectiveness of our algorithms at capturing the system's behavior due to submergence and emergence.  As the field shears, the boundaries between the three subregions have a tendency to discontinuously jump, resulting in large variations in the flux in each region, accounted for by the boundary algorithm.  However, the combined flux of the three subregions, shown in black, instead displays a fairly steady decrease: this whole region is submerging.  The blue line shows our reconstruction of this submergence using the source--change matrix at each time.  Each pole has an initial flux value, to which we add the elements of $\Delta^i\Matrix{S}$, so that each pole's flux at time $i$ is given by $\psi_j^i = \psi_j^0 + \sum_{l=0}^{i-1}\sum_k \Delta^{l}\Matrix{S}_{j,k}$.  This measure completely discounts the (artificial) boundary shifts between photospherically adjacent regions.  We then sum the flux in each pole thus reconstructed, and this is our estimate for the submergence of diffuse Western region.  We note that the blue and black lines follow each other reasonably well, indicating that our algorithms accurately capture the submerging trend of this region.
 
 \section{\label{sec:erg}Calculating the energy}
 \subsection{Topological definitions}
 In the second section of our analysis, we employ the Minimum Current Corona (MCC) model \citep{Longcope:2001,Longcope:1996} to use the amount of flux change in each domain to calculate the amount of free magnetic energy stored in our system as it evolves away from a potential configuration.  To that end, we must define a set of topological elements.  These definitions, briefly summarized below, are described more fully in \citet{LK:2002,Longcope:2005} and references therein.  The general idea is to model distributed photospheric sources as point sources.  The potential field generated by these sources will have null points, $\vect{x}_\alpha$, where the magnetic vector field vanishes: $\vect{B}(\vect{x}_\alpha) = 0$.  We may perform a linear expansion of the field about these points \citep[\S 2.4]{Parnell:1996,Longcope:2005}
 \begin{equation}
   \vect{B}(\vect{x}_\alpha+\delta\vect{x})\simeq \Matrix{J}^\alpha\cdot\delta\vect{x},
 \end{equation}
 where the Jacobian matrix $\Matrix{J}^\alpha_{ij}\equiv \partial B_i/\partial x_j$ is both symmetric and traceless because $\nabla\times\vect{B}=\nabla\cdot\vect{B}=0$.  As such, $\Matrix{J}^\alpha$ has three orthogonal eigenvectors with three real eigenvalues: the eigenvectors of the two like--signed eigenvalues define a plane (the fan); the eigenvector of the opposite--signed eigenvalue defines a line orthogonal to this plane (the spine).  From this foundation the rest of the topological description follows.  We define a \emph{pole}, the photospheric point source, located at the flux--weighted centroid of a mask.  \emph{Nulls} are the zeros of the potential magnetic field, and \emph{spines}, the fieldlines connecting a pole to a null along the single eigenvector.  A \emph{fan} (synonym \emph{separatrix}) is the set of fieldlines ending in a null's like--signed eigenvector plane.  A \emph{domain} is a simply connected space filled with field lines connecting a given pair of poles.  Finally, a \emph{separator} is the field line connecting two nulls, formed at the intersection of their respective fan surfaces.  In a nonpotential field, the separator may broaden into a two dimensional ribbon.  Coronal current sheets form along the separators of the field.  

 We determine the locations of all nulls in two steps.  First, we find those laying in the photospheric plane via the Newton--Raphson method of \citet{Barnes:2005}.  Second, we check the Euler characteristics given in \citet{LK:2002}, which provide both 2D and 3D relations between the number of nulls of each type and the number of sources of each polarity.  When the 3D characteristic is not satisfied, at least one coronal null is missing.  When then supply a ``by--eye'' list of initial $(x,y,z > 0)$ guesses to the same Newton--Raphson root finding method until coronal nulls have been found and the Euler characteristics are satisfied.  Other more automated coronal null detection methods have been developed \citep{Barnes:2007}, but the ad--hoc method described here has worked well in this case.

 After determining the location of all nulls, we located separators via the method of \citet{Barnes:2005}, with one modification.  Those authors only treat photospheric nulls, whose separators lay within separatrices above the photosphere.  For coronal nulls, we search for separators in all $2\pi$ directions within the null's fan surface.
 
 Our model photosphere consists of a plane $(z=0)$ of isolated sources surround by regions where the normal component of the magnetic field is zero.  This boundary condition may be satisfied using the method of images, as in \citet{LK:2002}\S 6.  With the normal field reflectionally symmetric in $z$, we have that $B_z(x,y,-z) = -B_z(x,y,z)$.  Whenever we introduce current in the corona along a separator, we must introduce its reflection in the mirror corona to maintain this symmetry.  In this way, our separators form closed current loops.

  \begin{figure}[ht]
   \capstart
   \centering
   \includegraphics[width=0.75\textwidth]{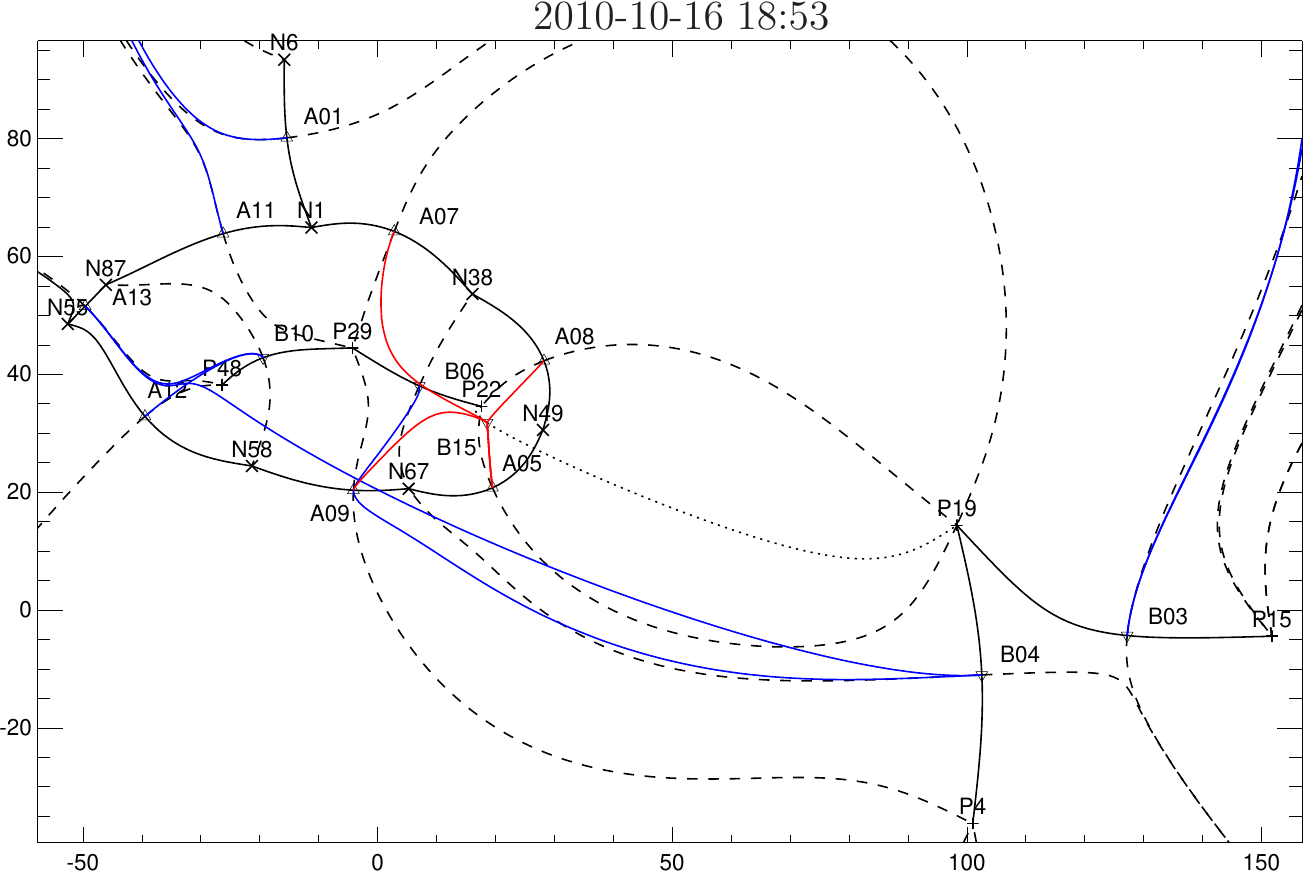}
   \caption[Topology of NOAA AR 11112]{\label{fig:topo}Topology of NOAA AR 11112 on the eve of the flare, depicted in a local tangent plane with coordinates.  The point of tangency is taken as the center of charge in the initial magnetogram, translated through solar rotation to the present time, and the axes are in Mm.  Pluses and crosses are positive and negative poles, respectively; triangles are positive ($\vartriangle$) and negative ($\triangledown$) nulls; solid black lines depict spines, dashed lines the trace of fans within the photosphere, and the dotted line is a coronal spine, attached to the coronal null B15.  Blue and red lines are projections of separators into the photosphere, with red lines those separators connected to the coronal null.}
 \end{figure}
  
  \subsection{Specifics of domain fluxes}
  Once we have created the set of matrices describing domain flux changes due to submergence or emergence, we can create similar matrices describing the flux change due to a changing potential field.  We calculate the potential field connectivity at each timestep via the Monte Carlo method of \citet[\S3.1]{Barnes:2005}.  This produces a set of connectivity matrices $\{\Matrix{P}\}$, where $\Matrix{P}^i_{j,k}$ is the flux connecting sources $j$ and $k$ at time $i$.  From these matrices we calculate the change in connectivity due to the changing potential field: 
  \begin{equation}
    \Delta^i\Matrix{P} \equiv \Matrix{P}^{i+1} - \Matrix{P}^i.
  \end{equation}
  Summation along rows of the connectivity matrix returns the total flux of the corresponding pole; eg. $\psi_j^i = \sum_k\Matrix{P}_{j,k}^i$.  We define $\Delta^i\Matrix{P}$ to be antisymmetric, as with the matrices $\Delta^i\Matrix{S}$ and $\Delta^i\Matrix{M}$ before.
  
  The MCC model derives an energy from the discrepency between the actual domain fluxes $\Matrix{F}^i$ at time $t_i$ and the potential fluxes then, $\Matrix{P}^i$.  In the absence of flux emergence or submergence, the actual fluxes are fixed using the potential field domain fluxes at some initial time: $\Matrix{F}^i = \Matrix{P}^0$, as in \citet{Kazachenko:2010}.  Any flux emergence or submergence through the photosphere, quantified as $\Delta^i\Matrix{S}$ in the previous section, modifies these actual domain fluxes, so that 
  \begin{equation}
    \label{eq:freal}
    \Matrix{F}^i = \Matrix{P}^0 + \sum_{j = 0}^{i-1}\Delta^i\Matrix{S}.
  \end{equation}
  
  Because we assume a field, initially potential, whose fluxes are fixed under future evolution, we determine how far removed the field is from a potential field configuration by answering the question, ``What flux must be added to a potential field domain $D$ at time $0$ to get a potential field domain at time $i$?''  This question is answered by
  \begin{equation}
    \label{eq:fquest}
    \Matrix{P}_D^i = \Matrix{P}_D^0 + \sum_{j=0}^{i-1}\Delta^j \Matrix{S}_D + \sum_{j=0}^{i-1} \Delta^j \Matrix{R}_D.
  \end{equation}
  On the right hand side, the first term is the potential field at the initial time.  The second is the total flux change through the photosphere.  The final term is flux change due to coronal redistribution, which must be achieved by modifying the connectivity matrix.  Together, the first two terms give the actual domain flux $\Matrix{F}^i_D$ at time $i$.  All flux changes from coronal reshuffling are then given by 
  \begin{equation}
    \label{eq:drx}
    \sum_{j=0}^{i-1} \Matrix{R}_D^j = \Matrix{P}_D^i - \Matrix{F}_D^i.
  \end{equation}
  Equation \eqref{eq:drx} holds for all domains, of course, and we represent this as a matrix equation by dropping the domain subscript $D$.  $\Delta^i\Matrix{P}$ is antisymmetric, and we have defined $\Delta^i\Matrix{S}$ to be antisymmetric, so $\Delta^i\Matrix{R}$ is also antisymmetric.  While antisymmetry of these matrices carries no physical information, we will later show that it does endow $\Delta^i\Matrix{R}$ with a nice mathematical property.
  
  Note that the difference in \eqref{eq:drx} between the actual domain fluxes and the potential field fluxes depends on the choice of the inital time $t_0$.  When dealing with actual data, we must pick an initial time to apply the MCC flux constraint when we believe the field is in a potential configuration in order to find a meaningful difference relative to a later potential configuration.  Ideally, we would track a single region from its inception up to an event which redistributes the coronal flux, as in a flare.  While NOAA AR 11112 does not fit this criteria, it likely matters little in this case.  As is apparent from our supporting media, and also as we will show below, NOAA AR 11112 has a very stable flux configuration for $\sim 40$hours prior to flux emergence.  This newly emerged flux, fixed in set domains as the photospheric field continues to evolve, rapidly diverges from a potential field configuration at later times.
  
  \subsection{\label{sec:seps}Separators, separator currents, and energy storage}
  So far this discussion only involves the sources themselves and their interconnections.  Let us now introduce a set of separators $\{\sigma\}$.  Separators are field lines that run from a null point of one type to a null of the opposite type.  In the MCC model, they are the sites of current sheets within the corona\citep{Longcope:2005,Priest:1996}.  Because current may only flow in loops, the separators themselves must close along some path in the mirror corona.  We can then speak of separator fluxes: the flux of domains linked by some separator $\sigma$ with its closure.  We will discuss closures and explain the concept of linking in detail below, but the general idea is that field lines in a linked domain cannot reconnect to another domain except by passing through the separator.  

  Let $\psi_\sigma^i$ be the flux linked by $\sigma$ at time $i$ in the actual field, and
  \begin{equation}
    \label{eq:linkpotl}
    \psi_\sigma^{(v)i} = \sum_D\Matrix{P}_D^i = \sum_{\{(j,k)\}}\Matrix{P}_{(j,k)\in D}^i
  \end{equation}
  be flux in the potential field domains $D=\{(j,k)\}$ linked by $\sigma$.  For a given closure, a separator always links the same set of domains throughout time.  The fluxes in those domains will generally change over time, however.  When a domain's flux goes to zero, the domain no longer exists.

  The flux--constrained--equilibrium (FCE) assumption of the MCC model states that domain fluxes are fixed as the field evolves.  Minimization of the field's energy subject to these constraints shows that the coronal field will be current--free except along the separator, where a current ribbon forms \citep{Longcope:2001}.  This occurs whenever $\psi_\sigma^i\ne\psi_\sigma^{(v)i}$.  We therefore define
  \begin{gather}
    \label{eq:fce}
    \psi_\sigma^{(cr)i} \equiv \Delta\psi_\sigma^i = \psi_\sigma^i - \psi_\sigma^{(v)i} = \sum_D\Matrix{F}^i_D - \sum_D\Matrix{P}^i_D,
    \intertext{which, from equation \eqref{eq:drx}, is}
    \label{eq:totfcr} \Delta\psi^i_\sigma= - \sum_D\sum_{j=0}^{i-1}\Delta^j\Matrix{R}_D.
  \end{gather}
  The double sum over elements of the redistribution matrix $\Matrix{R}$ is the difference between the flux linked by the separator in the actual and potential fields at time $i$.  It is also the self--flux generated by current flowing along the separator, and therefore, the flux that must be added to the real field to relax the FCE constraint of MCC: $\psi_\sigma^{(cr)i} +\sum_{D,j}\Delta^j\Matrix{R}_D = 0$.
  
  Our method for calculating energies therefore rests on two assumptions.  The first is that at some initial time $0 < i$ we assume that the domain fluxes are given by the potential field's domain fluxes, $\psi^0_{(j,k)} = \Matrix{P}^0_{(j,k)}$, so that all the separator fluxes $ \psi_\sigma^{(cr)0} =0$.  The second assumption is that the evolution of the current ribbon's self--flux depends on the evolution of the both the actual and potential field configurations:
  \begin{align}
    \label{eq:dfcr}\Delta^i\psi_\sigma^{(cr)} = \Delta^i\bigl( \Delta\psi_\sigma^i  \bigr) & = \Delta\psi_\sigma^{i+1} - \Delta\psi_\sigma^{i}\\
    & = \Delta^i\psi_\sigma - \Delta^i\psi_\sigma^{(v)}\\
    & = -\sum_D \Delta^i \Matrix{R}_D.
  \end{align}
  Summed over time, this is just the restatement of \eqref{eq:fce} itself, namely that the current ribbon's flux at some time $i$ is the sum of all changes in each linked domain's flux over the life of the separator.
  
  When $\psi^{(cr)i}_\sigma$ is small we can estimate the properties of the current ribbon, as in \citet{LongcopeMagara:2004}.  For a separator of length $L$ carrying current $I$, they determine the self--flux to be
  \begin{equation}
    \label{eq:crpsi}
    \psi^{(cr)i}_\sigma = \Frac{I L}{4 \pi}\ln\Biggl(\Frac{e I^*}{\vert I\vert}\Biggr), 
  \end{equation}
  where $I^*$ is a measure of magnetic shear in the separator's vicinity.  From this the excess energy of the MCC field relative to the potential field, i.e. the free energy, is\citep{LongcopeMagara:2004,Longcope:2001}
  \begin{equation}
    \label{eq:wmcc}
    \Delta W_{MCC} = \Frac{1}{4\pi}\int_{\Psi_{potl}}^{\Psi}I d\Psi = \Frac{L I^2}{32 \pi^2}\ln\Bigl(\Frac{\sqrt{e} I^*}{\vert I\vert}\Bigr).
  \end{equation}
  Note that \eqref{eq:crpsi} and \eqref{eq:wmcc} only take into account energy due to the self inductance of each current loop.  In general, we must include the effect of mutual inductance, though for physically separated loops we expect self inductance to dominate the total energy.
  
  The FCE assumption \eqref{eq:fce} requires that $\psi_\sigma^i = \psi^{(v)i}_\sigma + \psi^{(cr)i}_\sigma$, which allows us to solve \eqref{eq:crpsi} for the current in the current ribbon as a function of $\Delta\psi_\sigma^i$:
  \begin{gather}
    \label{eq:cr}
    I(\Delta\psi_\sigma^i) = I^*\Lambda^{-1}(4\pi\Delta\psi_\sigma^i/LI^*)
  \end{gather}
  where $\Lambda^{-1}(x)$ is the inverse of the function $\Lambda(x)\equiv x\ln (e/\vert x\vert)$.  Equation \eqref{eq:cr} provides a method for determining the current residing in, and hence the energy stored by, each separator: it tells us the current required to change the domain flux enclosed by a separator from that of a potential field to some other value.  In the present case that value is given by the non--zero elements of $\Delta^i\Matrix{R}$.  A reverse, cumulative sum over the time index $i$ for a given flux domain $P_j$--$P_k$ gives the time--history of departure from a potential field configuration.
  
  \subsection{\label{sec:gauss}Using the Gauss Linking Number to find linked domains}
  As is implicit in \eqref{eq:totfcr}, each separator may link more than one domain: the appropriate $\psi^{(cr)i}_\sigma$ for a separator is the sum of all changes in domain fluxes for each domain linked by that separator. Since all field lines in a given domain are topologically equivalent\footnote{This is \emph{not} to say that all flux connecting two poles is equivalent.  A single pair of poles may have more than one distinct domain: see \citet{BeveridgeLongcope:2005,Parnell:2007}.  This situation may arise, for instance, when you have purely coronal domains.}, we may establish separator linkage using a single representative fieldline and the separator itself.  After each of these open curves is closed, their linkage is found using the Gauss linking number\citep{BergerField:1984}, $L_{12}$, for each field line--separator pair: 
  \begin{align}
    L_{12} = L_{21} = \Frac{1}{4\pi}\oint_{\ell_1}\oint_{\ell_2}\Frac{\vect{r}_1 - \vect{r}_2}{\vert \vect{r}_1 - \vect{r}_2\vert^3}\cdot(d\vect{r}_1\times d\vect{r}_2)
  \end{align}
  The linking number not only determines whether a separator links a field line ($L_{12} \ne 0$), but also the sense in which it does so ($L_{12} = \pm n,\ n$ an integer): to calculate the correct $\psi_\sigma^{(cr)i}$, you must sum over all linked domains, multiplied by their respective linking numbers.

  In order to use Gauss' linking formula, we must have two closed curves, whereas our curves (usually\footnote{It is possible for a separator to attach to a coronal null, in which case, in order to create the appropriate loop, you must follow a second separator from the coronal null back to the photosphere, as discussed below.}) begin and end at two separate sources (nulls for separators) in the photosphere, never penetrating beneath.  So, in order to apply the linking formula in this case we must add to each coronal curve some curve in the mirror corona to form a complete loop.  There are, in general, many ways to form the closure for each curve, and not all of them will lead to the same linking number.  One requirement is that the separator be closed \emph{above} the field line's closure: any closure below the field line will always give a linking number of zero.  Therefore, we have chosen to close the separator with a straight line between the footpoints in the photosphere, and the field lines with a rectangular path formed by following the footpoints down and connecting them with a straight line in the $z=-0.5$Mm plane.  Another method would be to close the separator, say, with its projection into the $z=0$ (photospheric) plane.  Either method is valid, so long as the same method is used for all field lines and separators.

  Another requirement for using the linking number to find linked domains is that we follow field lines and separators in a consistent direction.  In general, changing the direction of \emph{one} of the integrals takes $L_{12}\longrightarrow \bar{L}_{12} = -L_{12}$.  For definiteness, we always trace separators from negative (A--type) nulls to positive (B--type) nulls above the photosphere, and thus close separators from B$\to$A in the photosphere.
  
  Because the reconnection matrix $\Delta^i\Matrix{R}$ is antisymmetric, this same care is unnecessary when tracing the field lines, at least for the purpose of the $\psi_\sigma^{(cr)i}$ calculation.  Let us designate by $\Delta^i\bigl(\Delta\psi_{P1,N2}\bigr)$ the difference in flux--change for the real versus potential field between times $i$ and $i+1$ (as in \eqref{eq:dfcr}), when we trace a field line from P1 to N2 in the corona, and close the field line from N2$\to$P1 below the photosphere.  Then, 
  \begin{align}
    \Delta^i\bigl(\Delta\psi_{P1,N2}\bigr) & = \Delta^i\Matrix{R}_{1,2} \times L_{12}
    \intertext{Now trace the field line from N2$\to$P1 in the corona, with the appropriate closure for $z<0$.  We have}
    \Delta^i\bigl(\Delta\psi_{N2,P1}\bigr) & = \Delta^i\Matrix{R}_{2,1} \times \bar{L}_{12}\notag\\
    & = \bigl(-\Delta^i\Matrix{R}_{1,2} \bigr)\times \bigl(-L_{12}\bigr)\notag\\
    & = \Delta^i\bigl(\Delta\psi_{P1,N2}\bigr),
  \end{align}
  We have therefore been justified in simply writing $\Delta^i\psi_\sigma^{(cr)}$, with no designation of tracing a field line from positive to negative or negative to positive in the corona, because $\Delta^i\psi_\sigma^{(cr)}$ is just a summation of multiple such domains.  This holds provided we always trace the separator in the same direction.

  Once determined, the linking number for each separator/field line pair provides the correct method for calculating each separator's total $\psi_\sigma^{(cr)i}$, including all linked domains, and hence the energy stored by each separator.
  
  \subsection{Two complications}
\begin{figure}[ht]
   \capstart
    \begin{center}
      \includegraphics[width=\textwidth]{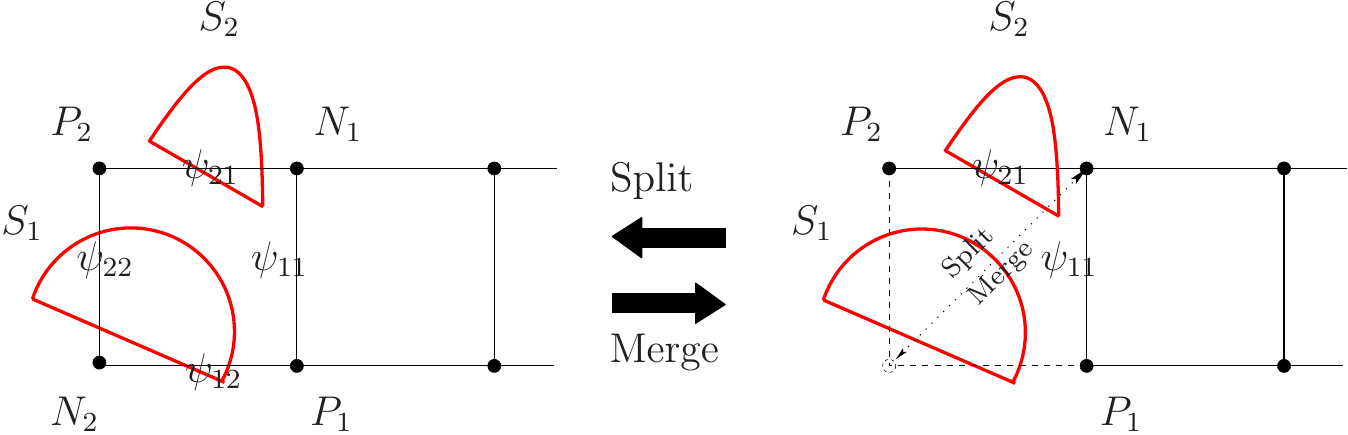}
      \caption[Topology of splitting poles]{\label{fig:constrict} Topological depiction of two poles either splitting or merging: $N_1\longleftrightarrow N_2$.  Dots depict the sources, lines the domains, and red arcs the separators.}
    \end{center}
  \end{figure}

  We must account for instances where two photospheric regions merge together or split apart, as illustrated in \figref{fig:constrict}.  Therein, we focus on 4 poles---$P_1, P_2, N_1, \& N_2$---with four flux domains---$\psi_{1,1}, \psi_{1,2}, \psi_{2,1}, \& \psi_{2,2}$---and two separators, $S_1$ and $S_2$, which link the domains in some fashion.  Suppose two of the poles are each connected to the same null via its  spines.  If the flux of one pole is zero before(after) a split(merger), then the null between them does not exist at that time, and the flux in the domains linking the sources goes to zero.  Now, if either $S_1$ or $S_2$ is attached to that null, then the separator will arise or cease to exist with that null, and the flux linked by it is appropriate.  Then again, if these separators are attached to some other set of nulls and happen to link these flux domains, our calculated linked flux is still appropriate.  Say flux domain $\psi_{2,2}$, for instance, either folds into or breaks out from domain $\psi_{2,1}$.  The actual break or merger constitutes a boundary shift ( with $N_2$ having zero flux either before or after the event) and so does not contribute to the flux linked by either separator.  Any additional flux change on top of this is either appropriately picked up by $S_1$ or $S_2$, but not both.

  A second complication arises when we encounter a coronal null, as we see in the topology of NOAA AR 11112 in \figref{fig:topo} near local tangent plane coordinates (20,30)Mm.  If a coronal null has any separators connected to it then it must have at least two: one carrying current up, the other down.  If many separators connect to the coronal null, then some must carry current up, and others down, but all balanced according to Kirchhoff's rules.  In order to calculate the flux within a separator--induced domain, we must close the separator somehow.  Above, we addressed this by prescribing a photospheric closure for separators.  Now, however, we must first stitch pairs of separators together in order to get back to the photosphere.  If we have $n$ separators connected to the coronal null, we require $n-1$ separator pairs, or ``isolating loops'' in the terminology of \citet{Longcope:2001}.  We might assume that the currents flowing along the $n-1$ loops would distribute to minimize the energy due to self and mutual inductances.  This problem remains to be solved, and for the moment we simply designate one separator linked to a coronal null as the ``shared'' separator.  We have found that the energies we calculate do not depend heavily on the choice of common leg, laying within $\approx 20\%$ of each other.

  As shown in \figref{fig:topo}, we have a single coronal null, B15, with spine lines leading to $P22$ and $P19$.  Four separators (in red) connect to the coronal null: from null A09 between $N58$ and $N67$, A05 between $N67$ and $N49$, A08 between $N49$ and $N38$, and A07 between $N38$ and $N1$.  Again, because all current flowing from the photospheric nulls up to the coronal null must be balanced by current flowing back down to the photosphere, we mush combine these four separators into three, all of which share one leg; it does not matter which one.  We have chosen A05--B15 leg to be common to all three separators.  As we will see below, the A05--B15--A08 separator generates the greatest amount of energy in this system.  This is expected, given the high horizontal shear in the photospheric field between $P22$ and $N49$, and the greatly flux--deficient domain between these two poles, relative to the potential field configuration at the time of the flare.
  
  \subsection{Summary}
  We may summarize the previous several sections in the following way.  In order to perform an energy calculation, we need a set of poles $\{P\}^{t_f}$ at a given time final time $t_f$; a set of nulls at the same time $\{N\}^{t_f}$; a set of separators at the same time $\{\sigma\}^{t_f}$; a time--history of the reconnective changes in domain fluxes from the final time, backwards to a previous time $t_i$ where we may assume the field was potential, $\{\Delta^i \Matrix{R}\}$; and a set of domains $\{D_\sigma\}^{t_f}$ linked by each separator.

  Having all these pieces, we then calculate the total domain flux difference in each domain at time $t_f$ relative to the initial presumed potential field.  For each separator, we find all domains linked by that separator.  Next, each linked separator's time--history of reconnective flux changes is summed.  This total flux--change for each linked domain is then multiplied by the Gauss Linking Number for that domain/separator pair.  Finally, the total domain changes are summed for each linked domain to fix the total flux difference of the separator at the final time $t_f$, relative to the potential field at the initial time $t_i$.  Mathematically, this is given by \eqref{eq:totfcr}, reproduced here:
  \begin{equation*}
    \psi_\sigma^{(cr)f} = -\sum_{D}\sum_{j=t_i}^{t_f-1}\Delta^j\Matrix{R}_D
  \end{equation*}
  This flux is generated by the current along the separator, \eqref{eq:crpsi}, which may be found by the inversion \eqref{eq:cr}, and plugged into the energy equation \eqref{eq:wmcc} to find the energy in excess of the potential field energy at time $t_f$ in the Minimum Current Corona model with the FCE constraint applied at time $t_i$.  This energy is a lower bound for any coronal energy model.

  \section{\label{sec:ergapp}Application of MCC energy calculation to NOAA AR 11112}
  By time of the flare (\figref{fig:newflux}) the new flux in NOAA AR 11112 has bubbled up and expanded within the interior of the old--flux ring.  The effect is especially pronounced between regions P22 and N49, the boundary of which has a very strong horizontal gradient in the vertical field.  Because this gradient is between old and new flux, prior to reconnection during the flare there is little flux in the domain connecting these two regions, and certainly far less than there would be in the potential configuration at that time.
  
  We expect one action of the flare will be to increase the flux within the P22--N49 domain.  Because its flux is conserved, any flux P22 adds to its N49 domain must come at the expense of its other domains.  The poles with which P22 shares these other domains---primarily N58, N55, and N87 with which P22 emerged---must in turn shuffle their flux into different domains.  In this way, a flare originating in one place propagates its effects throughout the entire magnetic structure until it establishes a new, lower energy equilibrium.

  \begin{figure}[ht]
   \capstart
    \begin{center}
      \includegraphics[width=\textwidth]{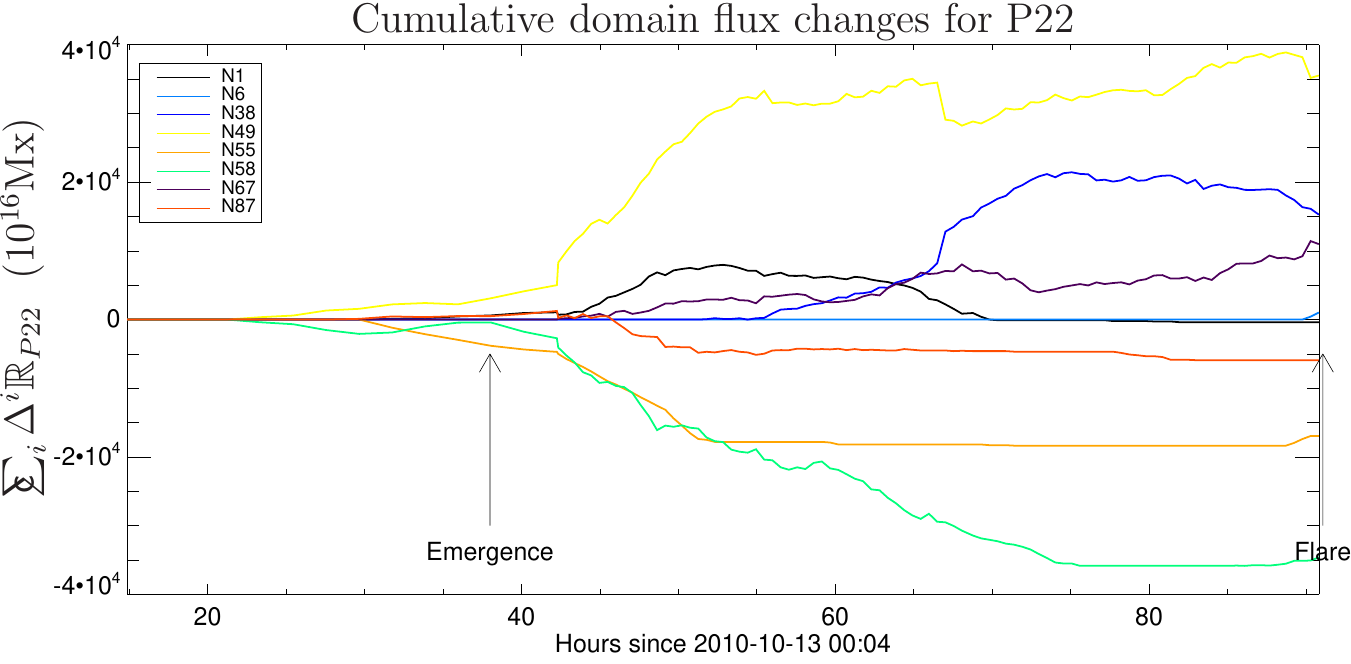}
      \caption[P22 domain flux changes]{\label{fig:p22rmat} Cumulative domain flux changes for various P22 domains, relative to the potential field at t=0.  Arrows indicate the onset of flux emergence and the UT 19:07 M2.9 flare.  There is some ``pre--emergence'' flux due to our 13 timestep smoothing function.}
    \end{center}
  \end{figure}

  We can get a picture of this process by looking at the cumulative domain flux changes for various P22 domains, shown in \figref{fig:p22rmat}.  Here, the flux at each timestep is given by the cumulative sum of all flux changes up to that time, $\sum_{j=0}^i(\Delta^j\Matrix{R}_{P22,*})$.  This figure shows by how much flux a domain must be changed to match the potential field configuration of the initial field, plus the contribution due to emergence.  

  P22 emerges in conjuction with N55, N58, and N87, thus all flux is assigned to domains P22--N55, P22-N58, and P22-N87.  The potential field, however, assigns less flux to these domains and more (i.e. some) to domains such as P22--N1 and P22--N49.  These latter domains are flux--deficient relative to the potential field, and appear above the zero line in \figref{fig:p22rmat}.  The former are flux--excessive, as a result of emergence, and appear below.

  Note that the potential field distributes the flux among domains differently at different times, and our analysis captures these changes.  In particular, P22 emerges to the East of N1, then migrates Westward.  As P22 moves past N1, the P22--N1 domain first becomes deficient, then more sufficient as the P22--N38 and P22--N67 domains become increasingly deficient.  The total flux in P22 only slowly increases throughout much this time (it is mostly emerged by 50hrs, as shown in \figref{fig:fluxplot}), so these later changes in the potential configuration are largely due to shear motion in the photosphere, rather than the initial emergence.

  AIA 171\AA\ images during the flare reveal how reconnection works to restore a flux balance closer to potential.  At successive times, we see brightenings of small loops progressing West to East within the newly emerged regions.  After this we see a series of loop brightenings connecting the new negative regions within the ring to the old diffuse positive region to the West.  We believe that this series of loop brightenings is a direct result of domain restructuring due to the emergence of the new flux and its highly nonpotential original field configuration.

  \begin{figure}[ht]
    \capstart
    \begin{center}
      \includegraphics[width=\textwidth]{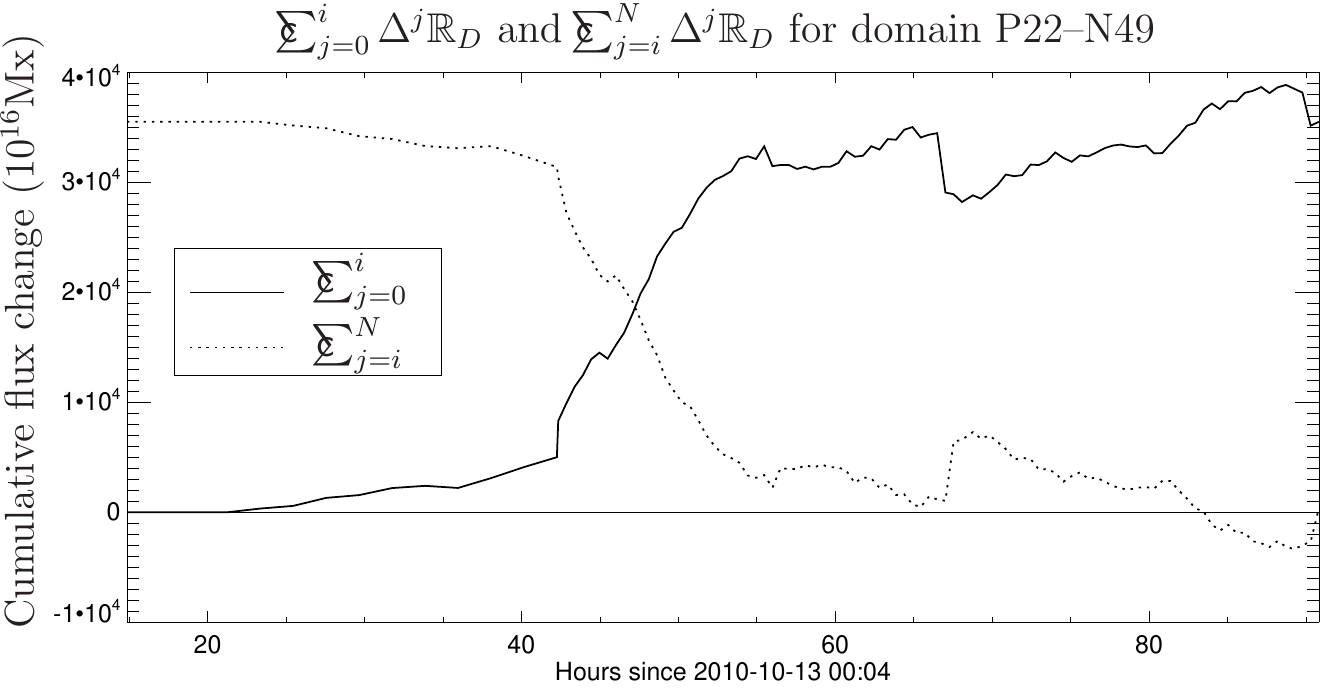}
      \caption[Two sums]{\label{fig:twosums}Two cumulative time--sums over $\Delta^i\Matrix{R}_{P22-N49}$.  The solid line sums from the initial time, 0, to time $i$ and is the flux required to change the domain at time $i$ to an assumed potential field at the initial time.  The dotted line sums from time $i$ to the final time, $N$, and is the flux required to change the final potential field configuration to a flux configuration fixed at time $i$.  The difference between the initial and final values is the same for both sums, and they are mirrors of each other about the average of this difference.}
    \end{center}
  \end{figure}
  
  The fluxes shown in \figref{fig:p22rmat} provide a good description for understanding the evolution of each domain, and directly follow the calculation outlined in \S~\ref{sec:seps}.  In this paper, however, we are interested in calculating the free energy due to a difference between domain fluxes fixed at some initial time and potential fluxes at the time of the flare.  For a completely new active region, we would assume all domain fluxes are zero except those which emerge together; in this case, our emergence occurs in a region of substantial old flux.  As such, our final time is fixed by the flare, and we must choose an appropriate time of initial constraint.  Combining equations \eqref{eq:freal} and \eqref{eq:fquest}, we find that $\Delta^i\Matrix{P}_D = \Delta^i\Matrix{F}_D + \Delta^i\Matrix{R}_D$, so that$\Delta^i\Matrix{R}_D$ is the flux that must be added to domain $D$ to account for change in addition to submergence and emergence, which is accounted for within $\Delta^i\Matrix{F}_D$.  \figref{fig:p22rmat} shows the cumulative flux $\sum_{j=0}^i\Delta^j\Matrix{R}_D$.  Wishing to vary the time of constraint, we calculate the reverse of this: $\sum_{j=i}^N\Delta^j\Matrix{R}_D$.  This is the amount of flux that must be added to the domain constrained at time $i$, $\Matrix{F}_D^i$, to match the potential field domain at time $N>i$.  As we will show below, this summation is basically constant prior to emergence: the old flux region is initially stable, and flux emergence drives a departure from this stable configuration.

  \figref{fig:twosums} gives an example of both the forward and reverse summations for domain P22--N49.  We set the fluxes for the energy calculation in this way because the assumption of the MCC model is that, barring reconnection, the domain fluxes are fixed, and prior to the flare there is no reconnection.  In order to compare with a potential field at the time of a flare, we determine our topology immediately before the flare.  This fixes all of the poles, nulls, and separators, and the domains linked by each separator.  The energy calculation then answers the question, ``\emph{Supposing the domain fluxes are fixed at time $i$, by how much must they be changed to match the potential field at the final time?}''  This difference for each domain linked by a separator then determines the amount of current along that separator, and hence the amount of energy stored by that separator.

   \begin{figure}[ht]
   \capstart
    \begin{center}
      \includegraphics[width=\textwidth]{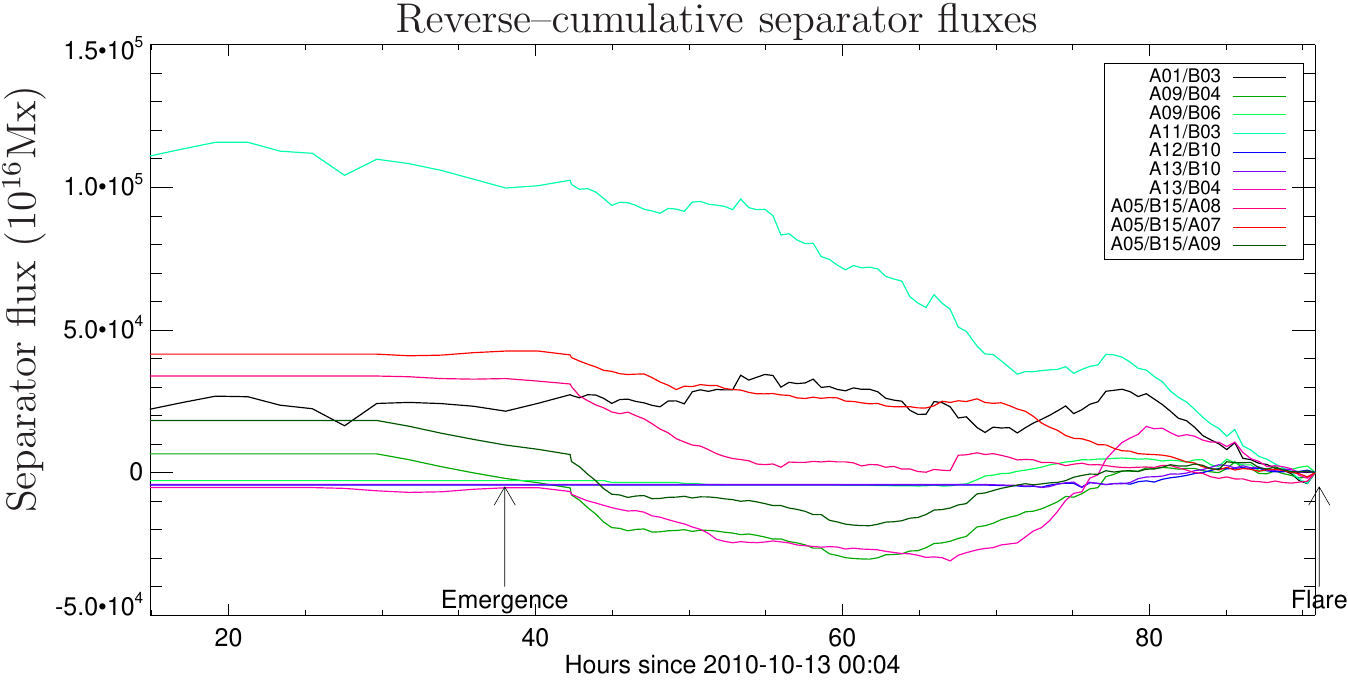}
      \caption[Separator fluxes]{\label{fig:sepflux} A reverse--cumulative sum of the total domain fluxes linked by each separator relative to the potential field configuration just prior to the flare, identified by nulls to which they attach.  Separators with three listed nulls are those which pass through the coronal null, B15.}
    \end{center}
  \end{figure}
  
   As mentioned previously, a single separator will generally link multiple domains.  For instance, one of the separators passing through the coronal null B15 in \figref{fig:topo}, connecting nulls A05--B15--A07, links three domains, N58--P22, N67--P22, and N38--P29, with Gauss Linking Numbers -1, -1, and +1, respectively.  The residual fluxes for two of these domains, N58--P22 and N67--P22, are shown in green and purple in \figref{fig:p22rmat}; the N38--P29 domain residual fluxes is similar, peaking at $\sim2.5\times 10^{20}$Mx around 90 hours.  This separator $\sigma$'s flux, including all three domains $\{D\}$ with linking numbers $L_D$, is then given at each time $i$ by
   \begin{equation}
     \psi_\sigma^{(cr)i} = \sum_{D}\sum_{j=i}^N\Delta^j\Matrix{R}_D\times L_{D}
   \end{equation}
   The sum for this separator is plotted in red in \figref{fig:sepflux}.  The curve shows that, as we look further back in time, the flux linked by this domain increasingly departs from the final potential field configuration, until around $t=43$ hours when all changes level off.  This is when region P22 first emerges.  Note that every separator shows relatively little change in the linked flux before this time.

\begin{figure}[ht]
    \capstart
    \begin{center}
      \includegraphics[width=\textwidth]{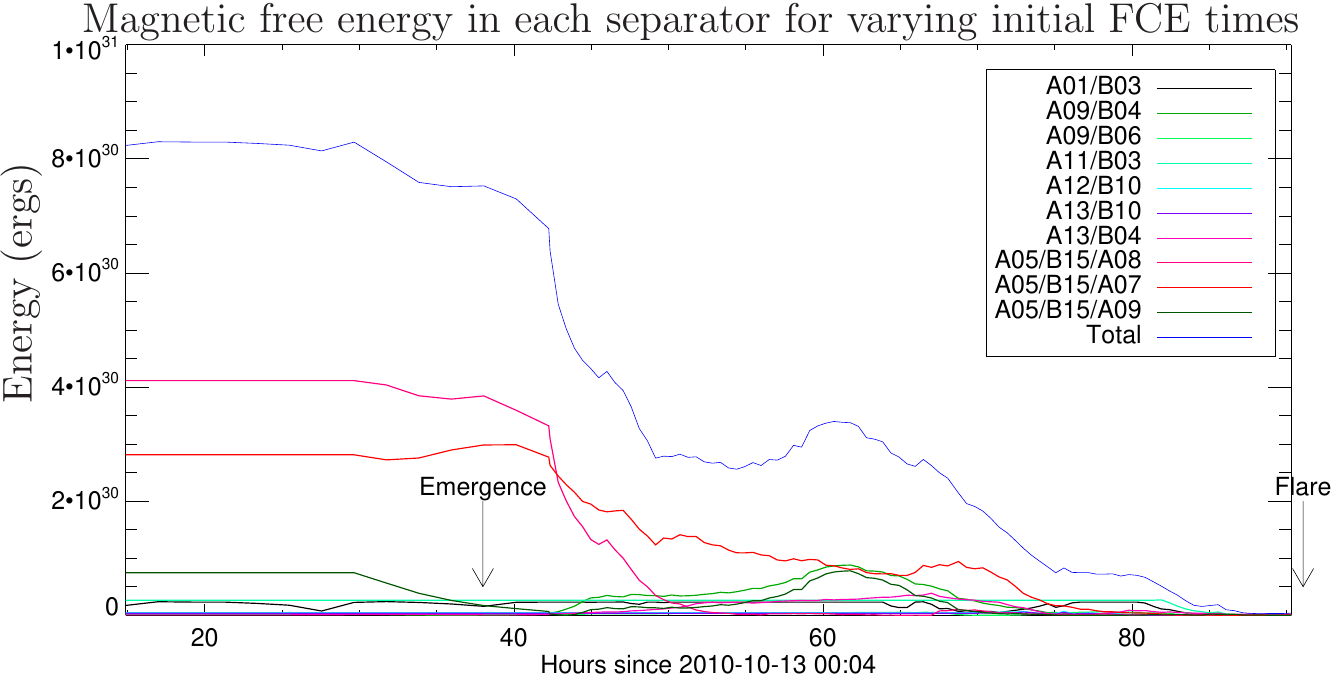}
      \caption[Energy storage in each separator]{\label{fig:energy} Free magnetic energy relative to a potential field at the final time, and applying the FCE assumption at successively earlier timesteps for each separator, and for the total (blue).  Our energy estimate is then the value of the initial timestep, $\approx 8.25\times 10^{30}$ergs.  Arrows indicates the onset of emergence and the time of the M2.9 flare.  Note that there is some ``pre--emergence'' due to our boxcar smoothing function.}
    \end{center}
  \end{figure}

   Based on \figref{fig:sepflux}, we might expect the A11/B03 separator to store the most energy, as it has the most linked flux.  However, the amount of current flowing in this separator is 1--2 orders of magnitude less than in the three separators connected to the coronal null: the longer a separator, the less current necessary to generate the required self--flux.  \figref{fig:energy} plots the magnetic free energy when applying the FCE assumption at varying timesteps.  We see that the energy of the three separators connected to the coronal null, shown in magenta, red, and dark green, completely dominate the magnetic free energy buildup.  The blue line shows the total for all separators.  We also see that, constraining the domain fluxes at any time before the onset of flux emergence, $t\approx 40$ hours, effects little change in the total energy buildup.  Essentially all of the flux difference relative to the final potential field configuration is due to the emergence itself.  

   The MCC yields a lower bound on the magnetic free energy of a system.  After accounting for all flux linked by all separators, and hence the energy stored in each separator, we conclude that NOAA AR 11112 stores a minimum of $8.25\times 10^{30}$ ergs of free magnetic energy over the 2--3 days leading up to the October 16\tothe{th}, 2010 GOES M2.9 flare.

  \section{\label{sec:conclusion}Conclusion}
  In this work we have extended the application of the Minimum Current Corona model to the estimation of energy stored along separators in the coronal field in active regions where flux submergence and emergence plays a significant role in the photospheric field's evolution.  This estimation operates in two distinct parts.  In the first, we track the evolution of the photospheric line--of--sight magnetic field using SDO/HMI magnetograms.  We found that the higher cadence and greater resolution of HMI, compared to previous analysis using full disk MDI data at a 96--minute cadence, greatly improved this portion of the analysis.  We found that a half hour cadence was sufficient to capture the evolution of this active region.  Analysis of this region began before the HMI 720 second averaged data were available.  Future investigations will use the high quality averaged dataset.

  The tracking information is stored in a mask array, which assigns unique labels to contiguous groups of data pixels, and matches these labels across all timesteps.  This framework accounts for all processes we see in the data: submergence and emergence, merging, splitting, braiding, and spinning.  In the present work we focus on submergence and emergence of magnetic flux.  All other processes are implicitly addressed in our work except for spinning.  \citet{Kazachenko:2010} have accounted for spinning explicitly in their work and found that its importance is case dependent.  We do not believe spinning plays a significant role in NOAA AR 11112.

  In the second part of our analysis we use the tracking information in the mask array to characterize the system wholly in terms of its topology.  Each region's total flux, together with its flux--weighted centroid, defines a magnetic pole.  A potential field extrapolation using the poles as sources, together with a mirror corona where $\vect{B}(x,y,z) = -\vect{B}(x,y,-z)$ yields all null points of the system.  The potential field extrapolation determines the locations of separators, which are where current sheets form due to free magnetic energy in the coronal field in the MCC model.  The amount of current flowing in each separator, and the amount of free energy in the field, is due directly to the nonpotentiality of each flux domain.

  A lower bound on the free magnetic energy is calculated from the separators and domain fluxes using the method of \citet{LongcopeMagara:2004}, who found that the MCC estimate may be less than other MHD free energy estimates by a factor of ten or more.  For NOAA AR 11112 we found a lower bound of $8.25 \times 10^{30}$ ergs.  This method requires that all domains linked by each separator are known.  This has been done by inspection in the past \citep{Kazachenko:2010}, but in the present work we have proposed an efficient automatic method using the Gauss Linking Number.  Provided the specification of some closure for both field lines and separators, and because domains are simply connected spaces, calculating the linking number for each separator and a single field line from each domain will find all linked domains for each separator, and the sense in which they are linked $(\pm 1)$.  It must be noted, however, that domains are not equivalent to connections: a single pair of positive and negatives poles may be multiply connected, with different domains in a single connection distinguished by one or more separators.  We found no such instances of this in the case of NOAA AR 11112, but this scenario may be simply dealt with by tracing a large number of random field lines between each pair of multiply connected poles.  One may then calculate the Gauss Linking Number of each separator with the suite of field lines, and determine the proportion of field lines linked by each separator.  The linked flux may then be divided among the separators, accordingly.

  Our free energy estimation does have some limitations.  Most important, we can only calculate energy storage, not energy release, for instance, in a flare.  This is because we fix the domain fluxes at some point in the past and compare them to the potential field domain fluxes at some given time; however the flare is a coronal phenomenon, while the potential field configuration is derived purely from photospheric data.  Using coronal data to quantify how domain fluxes change during a flare would allow for an energy estimation before and after a flare, though we are aware of no method for doing so.

  Second, our energy calculation only takes into account the self--inductance of each current ribbon.  For a set of currents loops, the total energy is given by both self and mutual inductance terms.  When the loops are physically separated, the self inductance terms should dominate.  However, if some subset of loops share a common leg (as is the case when multiple separators are connected to a coronal null point), then mutual inductance will significantly contribute to the total energy and must be accounted for.  This work remains to be done.

  \acknowledgements
  Graham Barnes graciously provided code for producing the potential field connectivity matrices using a Monte Carlo algorithm with Bayesian estimates, as described in \cite{Barnes:2005}.  Development of the code was supported by the Air Force Office of Scientific Research under contract FA9550-06-C-0019.  We thank the referee for his/her thorough reading and insightful comments which helped improve the manuscript.  We also thank our Summer 2010 REU student Johanna Bridge for her work in assessing the performance of the automatic tracking algorithms.  This work was supported by NASA LWS.

\addcontentsline{toc}{section}{References}

 \end{document}